\documentclass[11pt]{article}

\usepackage{amsmath,amstext,amssymb}
\usepackage{subeqn,xspace}
\usepackage{a4wide,url}
\usepackage{hyperref}
\usepackage[perpage]{footmisc}

\makeatletter
\renewcommand\section{\@startsection {section}{1}{\z@}%
                                   {-3.5ex \@plus -1ex \@minus -.2ex}%
                                   {2.3ex \@plus.2ex}%
                                   {\normalfont\large\bfseries}}
\renewcommand\subsection{\@startsection{subsection}{2}{\z@}%
                                     {-3.25ex\@plus -1ex \@minus -.2ex}%
                                     {1.5ex \@plus .2ex}%
                                     {\normalfont\normalsize\bfseries}}

\makeatother

\usepackage{theorem}
\newtheorem{theorem}{Theorem}[section]
\newtheorem{proposition}[theorem]{Proposition}
\newtheorem{lemma}[theorem]{Lemma}
\newtheorem{corollary}[theorem]{Corollary}

\newtheorem{definition}[theorem]{Definition}

{\theorembodyfont{\rmfamily} \newtheorem{remark}[theorem]{Remark}}

\newcommand{\Proof}{\noindent{\bfseries Proof.}\xspace}
\newcommand{\namedProof}[1]{\noindent{\bfseries Proof of #1.}\xspace}

\def\K{{\sf{K}}}

\newcommand{\ZZ}{{\mathbb Z}}
\newcommand{\NN}{{\mathbb N}}

\newcommand{\Endproof}{\noindent \hfill{\scriptsize $\square$}\vspace*{0.3cm}}

\newcommand{\softO}[1]{O{\tilde{~}}#1} 
\newcommand{\psoftO}[1]{O{\tilde{~}}(#1)} 

\def\MM{{\sf MM}}
\def\MMrecgjv{{\sf MM'}}
\def\MMrecas{\overline{{\sf MM}}}
\def\PM{{\sf M}}
\def\Pgcd{{\sf B}}
\def\ro{r_0}

\newif\ifshortversion

\shortversionfalse

\begin{document}

\title{{\normalsize \bf COMPUTING THE RANK AND A SMALL NULLSPACE BASIS\\
OF A POLYNOMIAL MATRIX}} 

\author{\normalsize 
Arne Storjohann\footnote{School of Computer Science, University of
  Waterloo, Waterloo, Ontario N2L~3G1 Canada
\hspace*{0.8cm}{\tt {http://www.scg.uwaterloo.ca/\~{}astorjoh}}}\,
and Gilles Villard\footnote{CNRS, LIP, \'Ecole Normale Sup\'erieure de Lyon 
46, All\'ee d'Italie, 69364 Lyon Cedex 07, France
\hspace*{0.8cm}{\tt {http://perso.ens-lyon.fr/gilles.villard}}}
}

\date{}

\maketitle

\begin{abstract}
We reduce the problem of computing the rank and a nullspace basis of
a univariate polynomial matrix to
polynomial matrix multiplication. For an input  $n \times n$ matrix of
degree~$d$ over a field $\K$ we give a rank and
nullspace
algorithm using 
about the same number of operations as for multiplying two matrices
of dimension~$n$ and degree~$d$.
If the latter multiplication is done in 
$\MM(n,d)=\psoftO{n^{\omega}d}$ operations,
with $\omega$ the exponent of matrix multiplication over $\K$,
then the algorithm 
uses $\psoftO{\MM(n,d)}$ operations in~$\K$. 
For $m\times n$ matrices of rank $r$
and degree $d$, 
the cost expression is $\psoftO{nmr ^{\omega -2}d}$. The soft-O
notation $\softO{}$ indicates some missing logarithmic factors.
The method is randomized with Las Vegas certification. 
We achieve our results in part through a combination of matrix
Hensel high-order lifting and matrix minimal fraction
reconstruction, and through the computation of 
minimal or small degree vectors in the 
nullspace seen as a $\K[x]$-module.
\end{abstract}

\def\thefootnote{}
\footnotetext[1]{
\newcount\hour\newcount\minutes\newcount\temp
\hour=\time \divide\hour by 60 \temp=\hour \multiply\temp by 60
\minutes=\time \advance\minutes by-\temp
{\hfill Text also available as Research Report 2005-3, Laboratoire LIP}

{\hfill ENSL, 46 All\'ee d'Italie, 69364 Lyon Cedex 07, France}

{\hfill
\url{http://www.ens-lyon.fr/LIP/Pub/Rapports/RR/RR2005/RR2005-03.pdf}
}
}

\def\thefootnote{\fnsymbol{footnote}}

\section{Introduction}\label{sec:intro}

Two $n \times n$ univariate polynomial matrices over a field~$\K$, whose entries have degree
$d$ at most, can be multiplied in
$\MM(n,d)=\psoftO{n^{\omega}d}$ operations
in~$\K$~\cite{BoSc04,CaKa91}
where $\omega$ is the exponent of matrix multiplication over $\K$~\cite[Chapter\,15]{BCS97}.
For $M \in \K[x] ^{n \times n}$ of degree $d$ we
propose an algorithm that uses about the same number of 
operations for computing the rank~$r$ of~$M$, and $n-r$ linearly
independent 
vectors $N_i$ in $\K[x] ^n$ such that $N_iM=0$, $1 \leq i \leq n-r$.
The cost of the algorithm is
$\psoftO{\MM(n,d)}=\psoftO{n^{\omega}d}$ operations in $\K$.
If $M$ is $m \times n$ of rank $r$,
a more precise and 
rank-sensitive expression of the cost is $\psoftO{nmr ^{\omega
  -2}d}$ (see Theorem~\ref{thetheo}).
The soft-O notation $\softO{}$ indicates 
missing logarithmic factors $\alpha (\log n)^{\beta} 
(\log d)^{\gamma}$ for three positive real constants $\alpha, \beta,
\gamma$. We mention previous works on the subject in Section~\ref{sec:previous}.
Our main idea is to combine matrix lifting
techniques~\cite{Sto02,Sto03}, minimal bases computation  and 
matrix fraction reconstruction~\cite{BeLa94,GJV03-2,JeVi04},
together with a degree\,/\,dimension compromise for keeping the cost
of the computation as low as possible. Within the target complexity,
lifting used alone only allows to
obtain few vectors of large degrees,  
while minimal bases used alone only leads to an incomplete set of vectors of small degrees.

Our study extends the knowledge of the interaction between matrix
multiplication and other basic linear algebra problems on matrices
over $\K[x]$. 
Indeed, the interaction 
is quite well known for 
linear algebra over a field.
For instance we refer to the survey~\cite[Chapter\,16]{BCS97}
for a list of problems on matrices in $\K ^{n \times n}$ that can be solved  
in $O(n^{\omega})$ or $\psoftO{n^{\omega}}$ operations in $\K$.
Only recent results give an analogous view (although incomplete) 
of the situation for polynomial
matrices. It is known that 
the following problems can be solved with $\psoftO{\MM(n,d)}$ operations:
{\em linear system solution}, {\em determinant}, {\em order~$d$
  approximants}, {\em Smith normal
form}, and, for a non-singular matrix, {\em column
reduction}~\cite{GJV03-2,Sto02,Sto03}. 
It is possible to compute
the {\em inverse} of a generic matrix in essentially optimal time
$\psoftO{n^{3}d}$~\cite{JeVi04}. We may also consider the problem 
of computing the Frobenius normal form, thus in particular the
characteristic polynomial,
 of a square matrix. It does not seem to be
 known how to calculate the form in time $\psoftO{\MM(n,d)}$.
The best known estimate $\psoftO{n^{2.7}d}$ is given
in~\cite{KaVi04-2} (see also~\cite{Kal92}) with $\omega=2.376$~\cite{CoWi90}.

Hence, we augment the above list of problems solved in 
$\psoftO{\MM(n,d)}$
with the {\em certified computation of the
rank} 
and a {\em nullspace basis}. This improvement is made
possible by combining in a new way the key ideas
 of~\cite{GJV03-2,JeVi04,Sto02}.
For the rank, the target complexity $\psoftO{\MM(n,d)}$ was 
only attainable by a Monte Carlo (non-certified) approach consisting
in computing the rank of $M(x_0)$ for $x_0$ a random
value in~$\K$ (see Lemma~\ref{lem:Mnr}). 
In obtaining a certified value of the rank and  a  nullspace 
basis within the target complexity, 
a difficulty is related to the output size. 
For $M \in \K[x]^{2n \times n}$ of degree $d$ and rank $n$, Gaussian
elimination (fraction free or using evaluation\,/\,interpolation)
leads to a basis of $n$ vectors  of degrees $nd$ in $\K[x]^{2n}$ in
the worst-case,
hence to an output size in $\Theta(n^3d)$. A complexity in 
 $\psoftO{n^{\omega}d}$ must therefore rely on a
 different strategy.

We propose a sort of elimination scheme based on {\em
  minimal polynomial bases}. A minimal basis  of the nullspace as
$\K[x]$-module is a basis with lowest possible degrees 
(all necessary definitions are given
in Section~\ref{sec:prelimin}). For $M \in \K[x]^{2n \times n}$ as
above,
the total size of a minimal basis of the nullspace 
is in $O(n^2d)$ (see Theorem~\ref{theo:transfer}).
However, it is not known how to reduce the problem of computing such
a basis to that of polynomial matrix multiplication. 
In the same context, minimal bases 
have been already used for computing the inverse of
a polynomial matrix in~\cite{JeVi04}, but only the generic case
has been solved. Indeed, for a generic $M \in \K[x]^{2n \times n}$, the
degrees in a minimal basis of the nullspace 
are all equal to the input degree $d$, and somehow, a basis is easy to
compute in   $\psoftO{\MM(n,d)}$ operations~\cite[Section\,4]{JeVi04}.
In the general case, the vector degrees in a minimal basis may be
unbalanced, they range between~$0$ to~$nd$. Known methods whose
cost is essentially driven by the highest degree do not seem 
to allow our objective.

Our solution presented in Section~\ref{sec:wholebasis} is to
slightly relax the problem, and to compute
a small degree---rather than minimal---nullspace 
basis in a logarithmic number of steps. 
We rely on the fact that even in the unbalanced degree case, the sum of the
degrees remains bounded by $nd$ (Theorem~\ref{theo:transfer}). Intuitively, at step\,$k$ for $1
\leq k \leq \log _2 n$, we
compute about $n/2^k$ vectors of degrees less than $2^kd$.
Algorithm {\sf Nullspace}$(M)$
in Section~\ref{sec:wholebasis} (whole nullspace) calls  at most $\log _2 n$ times
Algorithm {\sf Nullspace minimal vectors}$(M,\delta)$ of Section~\ref{sec:smalldeg}
(nullspace vectors of bounded degree~$\delta$) 
with increasing degree thresholds~$\delta$. To keep the cost as low as
possible,
the degree increase requires to reduce the dimensions of involved
matrices in the same proportion.
We refer to an analogous degree\,/\,dimension  compromise 
in~\cite[Section\,17]{Sto03} for computing the Smith normal form,
and in~\cite[Section\,2]{JeVi04} for inversion.

For a general view of the process, including successive compressions
of the
problem into smaller problems for reducing dimensions,
consider 
\begin{equation} \label{eq:geneq}
M = \left[ \begin{array}{c} A \\ B \end{array} \right] \in
\K[x]^{m \times n} 
\end{equation}
with $A$ square and non-singular.
The rows of the matrix $[BA ^{-1}~~-I_{m-n}]$ 
give a basis of the nullspace of $M$. However,
as noticed previously a direct calculation of 
$B A ^{-1}$ would be too expensive. 
Now, note that if $[B A ^{-1} ~~-I_{m-n}] = S ^{-1} N$, for $S$ and $N$ two
appropriate polynomial matrices, then 
the rows of $S[BA ^{-1}~~-I_{m-n}] = N$ are also in the nullspace.
A key observation, see Section~\ref{sec:MFD}, is that 
considering a polynomial matrix $N$ instead of $[B A ^{-1} ~~-I_{m-n}]$
 takes advantage of minimal bases properties and allows us
to manipulate smaller degrees.
 
In algorithm {\sf Nullspace} we proceed the following way.
We deal with a small number of submatrices of the initial input for reducing
the problem to
\begin{equation} \label{eq:geneqp}
M = \left[ \begin{array}{c} A \\ B \end{array} \right] \in
\K[x]^{(n+p) \times n}, ~ 1 \leq p \leq n, 
\end{equation}
and introduce appropriate ``compressing'' matrices $P \in \K[x] ^{n \times p}$
(successive choices of $p$ are guided by the compromise with the
degree). 
We start with a matrix lifting\,/\,fraction reconstruction phase.
We compute an expansion of $H=B A ^{-1}$ in $\K[[x]]^{p \times n}$ 
using~\cite{Sto02,Sto03} to sufficiently high order,
and ``compress'' it to $H_p=B A ^{-1}P \in \K[[x]]^{p \times p}$. 
A reconstruction phase~\cite{BeLa94,GJV03-2} (see
also the comments about coprime factorization in
Section~\ref{sec:previous}) then gives 
\begin{equation} \label{eq:reconintro}
S ^{-1}N_p = B A ^{-1}P.
\end{equation}
We prove that ``good'' choices of $P$ imply that
$S$---denominator matrix for $B A ^{-1}P$---is also 
a denominator matrix for $B A ^{-1}$ (Proposition~\ref{prop:lrHp}) and that
vectors in the nullspace of $M$ can be recovered (Proposition~\ref{prop:HpH}).
Indeed, the computation of $S[H ~~ -I_{m-n}] \bmod x^{\delta +1}$
gives row vectors in the nullspace of degrees bounded by $\delta$ 
(Proposition~\ref{prop:pnullspace}).
For a candidate Monte Carlo value $\ro$ for the rank, 
in $\log _2 n$ steps of compression\,/\,uncompression (and choices
of~$\delta$ and~$p$) 
combined with matrix lifting\,/\,matrix fraction reconstruction,
we are able to compute candidate vectors for a nullspace basis. 
A final multiplication certifies that the rank is correct (i.e.,  $\ro=r$)
and that a nullspace has actually been computed.

Although for each degree threshold~$\delta$ we compute a minimal polynomial
basis, the compression strategy unfortunately does not lead to a minimal polynomial
basis for the whole nullspace. However, we prove especially
in Proposition~\ref{prop:nullspace}
 that vectors with reasonably 
small degrees are obtained.

Our algorithms are randomized of Las Vegas kind--- always correct,
probably fast. Randomization is essentially linked to the
compression stages where the matrices $P$ are chosen at random of
degree~$d$ in $\K[x]^{n \times p}$. We also use random matrices $Q$ 
over $\K$ for linear independence preconditioning~\cite{CEKSTV01-2},
or random evaluation points $x_0$ in $\K$.
Our results are proven for symbolic points $x_0$ and matrices~$P$
and $Q$.
By evaluation~\cite{DeMLi78,Zip79,Sch80},
the same results hold with high probability 
for random $x_0$, $P$ and $Q$ if~$\K$
has enough elements, see Remark~\ref{rem:genrand}.
The cost estimates might increase by  poly-logarithmic 
factors in the case of small fields (with the introduction of 
an algebraic extension).
We skip the details here, and refer for instance to the techniques used
in~\cite{CEKSTV01-2,KaKrSa90,KaSa91}
and to the references therein. 

We study the cost of the algorithms by bounding the number of field
operations in~$\K$ on an algebraic random access machine.
In~\cite{GJV03-2} and~\cite{Sto03}, {\em ad hoc} cost functions 
have been defined for matrix polynomial problems that can be reduced
recursively to matrix polynomial multiplication:
$$
\MMrecgjv(n,d) = \sum _{i=0}^{\log _2 d} 2^i \MM (n, 2^{-i}d) 
$$
and 
$$
\MMrecas(n,d) = \sum _{i=0}^{\log _2 n} 4^i \MM (2^{-i}n,
d) + n^2 (\log n) \Pgcd(d)
$$
where $\Pgcd (d)$ is the cost for solving the extended gcd problem
for two polynomial in~$\K[x]$ of degree bounded by~$d$.
If $\PM(d)$ is the number of
operations in~$\K$ required for multiplying two polynomials in $\K[x]$
of degree~$d$, the
Knuth~\cite{Knu70}\,/\,Sch\"onhage~\cite{Sch71} half-gcd algorithm 
allows $\Pgcd(d)=O(\PM (d) \log d)$.
For the scalar polynomial multiplication we take $\PM(d) = O(d \log
d \log \log d)$~\cite{CaKa91}.
The reader may refer to Chapters~8 and~11 in~\cite{vzGG99} for more
details and references about polynomial multiplication and gcd computation.

For simplifying the cost results in this paper we consider either that
\begin{equation} \label{eq:costMM1}
\MM(n,d) = O(n ^{\omega} \PM(d))
\end{equation}
using the algorithm of~\cite{CaKa91},
or, when the field~$\K$ has at least $2d+1$ elements~\cite{Bos03,BoSc04},
\begin{equation} \label{eq:costMM2}
\MM(n,d) = O(n ^{\omega}d + n^2 \PM(d)).
\end{equation}
Hence from~(\ref{eq:costMM1}) and~(\ref{eq:costMM2}) we assume that 
\begin{equation} \label{eq:MMrec}
\MMrecgjv(n,d) = O(\MM(n,d) \log d),~~ \MMrecas(n,d) =
O((\MM(n,d) +n^2 \Pgcd(d)) \log n).
\end{equation}
Note that if $\omega >2$ then $\overline{{\sf MM}}(n,d) = O(\MM(n,d)
+ n^2 \Pgcd(d)\log n)$.
If the assumption~(\ref{eq:MMrec}) is not made then some of our 
cost results that use $\MM(n,d)$ are not valid. However, we state our
algorithms 
in terms of polynomial matrix multiplication; 
precise complexity estimates in terms of the {\em ad hoc} cost functions
could be derived with some extra care.


\section{Previous works}\label{sec:previous}

The rank and a basis for the nullspace of a matrix $M \in \K[x] ^{m
  \times n}$ of degree~$d$ and rank~$r$ 
may be computed by fraction free Gaussian elimination 
in $\psoftO{nmr ^{\omega -1}d}$ operations in~$\K$~\cite[Chapter\,2]{Sto00}.
The same asymptotic estimate may also be obtained using
evaluation\,/ interpolation techniques such as Chinese
remaindering~\cite[Section\,5.5]{vzGG99}.

Therefore, compared to these classical approaches, we improve the
cost by a factor $n$ in the worst-case ($2n \times n$ full
column-rank matrix).

An elimination strategy specific to polynomial matrices is given
in~\cite{MuSt02-2}
that improves---asymptotically in the dimensions---on $\psoftO{nmr ^{\omega
  -1}d}$, 
and computes the
rank
by a deterministic algorithm in $O(nmrd^2)$ operations in~$\K$,
but how to incorporate matrix multiplication, and generalize the
approach to computing the nullspace, is not known.

An alternative to the ``matrix over the polynomials'' approach above
is to linearize the problem.
A first type of linearization is to 
consider a degree one matrix of
larger dimension with the same structural invariants (see
the definition of the Kronecker indices in
Section~\ref{sec:prelimin})~\cite{BvD88}.
A degree one matrix is a matrix pencil and an important literature
exists on the topic. A minimal nullspace basis of a pencil may be
computed through the calculation of the Kronecker canonical form. 
To our knowledge, the best known complexity for computing the 
Kronecker form of an $m \times n$ pencil is $O(m^2n)$~\cite{BvD88-2,MvDV94,OaVD97}.
Taking into account the dimension increase  
due to the linearization we may evaluate that computing a minimal basis 
of $M$ would cost $O((md)^2(nd))=O(m^2nd^3)$. This approach is superior to ours
concerning the quality of the output basis which is minimal. However,
it is 
unclear how it can lead to the 
reduction to
polynomial matrix multiplication that we establish.

A second alternative and different linearization of the problem is
to associate to $M$ a generalized Sylvester 
matrix (i.e., a block-Toeplitz matrix~\cite{BKAK78}) or
another type of resultant.
This has been heavily used for control theory problems and in
linear algebra. A polynomial vector of degree $\delta$ in the nullspace of $M$ may be
obtained from the nullspace of a block-Toeplitz of dimension about
$n\delta$.
This leads to costs too high by a factor of $n$ when the degrees in a
minimal nullspace basis are unbalanced. We are not aware of an
approach based on successive compression here that would allow to  
save a factor $n$ and to introduce polynomial matrix
multiplication.

These two types 
of linearization correspond 
to two main approaches---based on state-space realizations or on
resultants--- for the problem 
of {\em coprime matrix fraction description}
or {\em coprime factorization}~\cite[Chapter\,6]{Kai80}.
We see from~(\ref{eq:reconintro}) that we will use a solution to the
latter problem a logarithmic number of times on the compressed matrices.
If all matrices involved are of degree $d$, then we use the
$\sigma$-basis algorithm of~\cite{BeLa94}, and the corresponding
reduction to polynomial matrix multiplication of~\cite{GJV03-2}.
A solution of the coprime factorization in case of unbalanced degree
is, in a way similar to the block-Toeplitz approach, 
is faced with the question of saving a factor $n$ in
the cost. Known algorithms seem to have a cost driven  only by the highest 
degree in the factorization, rather than by the sum of the
involved degrees
as we propose.  

Our work is a derivation of an elimination scheme using minimal
bases directly on polynomial matrices. Our
compression\,/\,uncompression strategy can be compared to the
techniques used for the staircase algorithm of~\cite{BvD88-2,OaVD97}
for preserving a special structure.
We somehow generalize the latter to the case of polynomial matrices 
for reducing the matrix description problem with input $B A ^{-1}$ 
to the polynomial matrix multiplication.


\section{Preliminaries for polynomial matrices}\label{sec:prelimin}

We give here some definitions and results about minimal
bases~\cite{For75} and matrix fraction descriptions that will be used in the rest
of the paper. For a comprehensive treatment we refer
to~\cite[Chapter\,6]{Kai80}. For a matrix $M \in \K[x]^{m \times n}$
of rank~$r$ and degree~$d$, we call (left) nullspace the $\K(x)$-vector space
of vectors $v \in \K(x)^m$ such that $vM=0$. 
We will compute a basis of that space. 
The basis will be given by
$m-r$ linearly independent polynomial vectors, and is related to the
notion of minimal basis of the nullspace seen as a $\K[x]$-module.

\begin{definition} \label{def:minbasis}
A basis $N_1, \ldots , N_{m-r} \in \K[x] ^m$ with degrees $\delta
_1 \leq \ldots \leq \delta _{m-r}$ of the nullspace 
of $M$ seen as a $\K[x]$-module is called a {\em minimal basis} if any other
nullspace basis with degrees $\delta'
_1 \leq \ldots \leq \delta' _{m-r}$ satisfies 
$\delta ' _i \geq \delta _i$ for $1 \leq i \leq m-r$.
\end{definition}

In the rest of the text, ``basis'' will usually refer to the
vector space while ``minimal basis'' will refer to the module.
The degrees $\delta
_1, \ldots , \delta _{m-r}$
are structural invariants of the nullspace.
They are called the minimal indices of the nullspace basis.
The minimal indices of a nullspace  basis  of $M$ are 
called the (left) {\em Kronecker indices} of~$M$.
A polynomial matrix $M \in \K[x]^{m \times n}$ is called {\em row-reduced} if its
leading row coefficient matrix has full rank. It is called 
{\em irreducible} if its
rank is full for all (finite) values of $x$
(i.e., $I_m$ is contained in the set of $\K[x]$-linear combinations 
of columns of $M$).
These two definitions are used for characterizing
minimal bases; we refer to~\cite[Theorem\,6.5-10]{Kai80} for the
proof of the following.

\begin{theorem} \label{theo:rowred}
The rows of $N \in \K [x]^{m-r}$, such that $NM=0$, form a minimal basis of the
nullspace of $M$ if and only if $N$ is row-reduced and irreducible.
\end{theorem}

A key point for keeping the cost of the computation low is the
degree transfer between $M$ and a minimal nullspace basis
$N$. 
The McMillan  degree of~$M$ of rank~$r$ is the maximum of the
degrees of the determinants of $r \times r$ submatrices of
$M$~\cite[Exercise\,6.5-9]{Kai80}.

\begin{theorem} \label{theo:transfer}
\ifshortversion
\else
The Kronecker indices and the McMillan degree of $M$ satisfy 
\begin{equation} \label{eq:kromcmillan}
\sum _{i=1}^{m-r} \delta _i \leq \text{\rm McMillan-deg\,} M 
\end{equation}
with equality if $M$ is irreducible.
\fi
\end{theorem}
\ifshortversion
\else
\ifshortversion
\vspace*{0.4cm}
\namedProof{Theorem~\ref{theo:transfer} page \pageref{theo:transfer}}
\else
\Proof
\fi
Let $I$ and $J$ be row and column index sets such that 
the McMillan degree of the submatrix  $M_{I,J}$ of $M$
is equal to the one of $M$.
By considering $M_{\cdot,J}$ we reduce ourselves to the full
column-rank case. 
Define $I_c = \{1, \ldots, m\} \setminus I$, and
the corresponding submatrices $A=M_{I,J}$ and $B=M_{I_c,J}$ of $M$.
By unimodular column reduction we may assume that $M_{\cdot,J}$ is
column reduced,
since $A$ carries the McMillan degree, $BA ^{-1}$ is proper.
A minimal basis $N$ gives the corresponding matrices $C=N_{\cdot,I}$
and $D=N_{\cdot,I_c}$
such that $CA+DB=0$. The matrix $D$ cannot be singular otherwise
there would exist a vector $u\neq0$ such that $uD=0$ and $uC\neq 0$
(the latter since $N$ is non-singular). This would give 
a non-zero vector $u$ such that $uA=0$ which is not possible.
Hence, $D^{-1}C=BA ^{-1}$. If $M$ is irreducible, then since $N$ is
irreducible by definition, both latter fractions are irreducible.
By~\cite[Theorem\,6.5-1]{Kai80}, $\deg \det D = \text{\rm
  McMillan-deg\,} M$.
Therefore, using the fact that $D^{-1}C$ is proper we know that $\deg \det D =
\sum _{i=1}^{m-r} \delta _i$, and the Theorem is
established.
When $M$ is not irreducible the same reasoning applies with $\deg
\det D \leq \text{\rm
  McMillan-deg\,} M$.
\Endproof
\fi

As dicussed in the introduction, Gaussian elimination is far too
pessimistic when it results in a nullspace basis with degree sum
in $\Theta(n^3d)$. Theorem~\ref{theo:transfer} shows that 
there exist minimal bases with degree sum in $O(n^2d)$ whose
computation should be cheaper.  

We will use minimal bases in relation with left or right 
matrix fraction descriptions.
A left fraction description $S ^{-1} N$ is irreducible (or coprime)
if any non-singular polynomial matrix 
and left common divisor $U$ of $S$ and $N$ (i.e.,
$S=US'$ and $N=UN'$ for polynomial matrices $S'$ and $N'$) is unimodular.
An analogous definition holds on the right.

\begin{lemma} \label{lem:fracbasisred}
The rows of $N=[N_p ~~S]$, such that $NM=0$, with $S$ non-singular form a basis for the
nullspace as a $\K[x]$-module if
and only if
$S^{-1}N_p$ is irreducible.  
\end{lemma}
\ifshortversion
\else
\ifshortversion
\vspace*{0.4cm}
\namedProof{Lemma~\ref{lem:fracbasisred} page \pageref{lem:fracbasisred}}
\else
\Proof
\fi
We have that $N$ is a basis if and only if it is irreducible, which
in turn is equivalent to the fact that $S$ and $N_p$ are
coprime~\cite[Lemma\,6.3-6]{Kai80}.
\Endproof
\fi

For a rational matrix ${\mathcal H}$ we may define the $\K[x]$-module
${\mathcal P}_{\mathcal H}$ 
of polynomial vectors $u$ such that $u{\mathcal H}$ is polynomial.  
We will use the following.

\begin{lemma} \label{lem:muldenom}
$S^{-1}N={\mathcal H}$ is a coprime matrix description of ${\mathcal
  H}$ 
if and only if the rows of $S$ form a basis of  
${\mathcal P}_{\mathcal H}$.
\end{lemma}
\ifshortversion
\else
\ifshortversion
\vspace*{0.4cm}
\namedProof{Lemma~\ref{lem:muldenom} page \pageref{lem:muldenom}}
\else
\Proof
\fi
Consider $T$ non-singular whose rows are in~${\mathcal P}_{\mathcal H}$. Then 
for a polynomial matrix $M$ we have $T{\mathcal H}=TS^{-1}N=M$, hence
$S^{-1}N=T^{-1}M$. Since $S^{-1}N$ is coprime, $T$ is a left
multiple of $S$~\cite[Lemma\,6.5-5]{Kai80}. Conversely, 
if the rows of $S$ form a basis of ${\mathcal P}_{\mathcal H}$, then $S^{-1}N$ 
is coprime. Otherwise, $S$ would be a multiple of $S_c$ for
$S_c^{-1}N_c$ coprime, which would contradict the basis property.
\Endproof
\fi

In Section~\ref{sec:smalldeg} we will focus on computing only
vectors of degrees bounded by a given $\delta$ in a nullspace
minimal basis. We define their number $\kappa 
=\max \{1 \leq i \leq m-r \text{~s.t.~} \delta _i
\leq \delta\}$ (the Kronecker indices are arranged in increasing order).
Corresponding vectors are called $\kappa$ {\em first minimal vectors} in
the nullspace.

\begin{remark}\label{rem:submodule}
We will also manipulate the module generated by $\kappa$ such vectors.
As in Theorem~\ref{theo:rowred}, a corresponding submatrix
$\tilde{N}$ with $\kappa$ rows of $N$ must be irreducible. As in
Lemma~\ref{lem:fracbasisred}, if 
$\tilde{N}=[\tilde{N}_p~~\tilde{S}]$ then $\tilde{N}_p$ and
$\tilde{S}$ have no left and non-singular common divisor
other than unimodular.
Since a minimal basis $N$ of the nullspace is row-reduced, by the 
predictable-degree property~\cite[Theorem\,6.3-13]{Kai80}, any
vector of degree less than~$\delta$ must be in the sub-module
generated by $\kappa$ minimal vectors. 
\end{remark}


\section{Matrix fraction descriptions for the nullspace} \label{sec:MFD}

Let us consider a matrix $M=[A ^T ~~ B^T]^T  \in  \K[x]^{(n+p)
  \times n}$ 
of degree $d$
as in~(\ref{eq:geneqp})  
with $A$  square $n \times n$ and invertible.
Our study here and in next section focuses on the case $p \leq n$
which is the heart of the method, and  
where all difficulties arise. The results here remain true but are
trivial for $p > n$ (see Remark~\ref{rem:bigp}).

The rows of ${\mathcal H} = [H~~-I_p] = [ BA^{-1} ~~-I_p]$ form a nullspace basis of $M$.
Hence, for $N$ a minimal nullspace basis, there exists 
a transformation $S$ in $\K(x) ^{p \times p}$ such that $S{\mathcal H}=N$.
With the special shape of ${\mathcal H}$ we deduce that $S$ is a polynomial matrix in $\K[x]^{p \times p}$ 
whose columns are the last $p$ columns of $N$. This leads to the following
left matrix fraction description of ${\mathcal H}$:
\begin{equation} \label{eq:H}
{\mathcal H} = [H~~ -I_p] = [ BA^{-1} ~~
-I_p] = S^{-1} N.
\end{equation}
The left fraction description $S^{-1} N$ must be irreducible
otherwise there would exist another description 
${\mathcal H}=(S')^{-1} N'$ with $N' \in \K[x]^{p \times (n+p)}$ having row
degrees lexicographically smaller than the
row degrees of $N$. Since $N'M=0$ this would contradict the
fact that $N$ is minimal.

For reducing the cost of our approach we will introduce a (random) column
compression $H_p$ of $H$ given by   
\begin{equation} \label{eq:Hp}
H_p = HP = B A ^{-1}P \in \K[x] ^{p \times p}
\end{equation}
with $P \in \K[x]^{n \times p}$.

In order to be appropriate for
computing the nullspace of $M$, $H_p$ must keep certain invariants
of $B A ^{-1}$. 
We establish in the rest of the section--- see
Proposition~\ref{prop:lrHp}---that 
there exists a $P$ such
that, on the left, the description $H_p=S^{-1} (NP)$ {\em remains irreducible}.
With the same $P$ we show the existence, on the right, of a
description whose {\em denominator matrix has relatively small degree}.
The existence of such a $P$ will ensure that the properties remains
true for a random compression.

\begin{lemma}\label{lem:realA}
Let $A$ be non-singular of degree less than 
$d$ and determinantal degree 
$\nu \neq 0$ in $\K[x] ^{n \times n}$.
Let $B$ be in $\K[x] ^{p \times n}$.
There exists a surjective function $\sigma: \K[x]^{n \times p} \rightarrow \K ^{\nu
  \times p}$, and two matrices $X \in \K^{p \times \nu}$ and 
$A_o \in \K^{\nu \times \nu}$, such that for any $P$ in $\K[x] ^{n \times p}$
\begin{equation}\label{eq:real}
H_p(x) = B(x)A(x) ^{-1} P(x) = Q(x) + X (x - A_o) ^{-1} \sigma (P),
\end{equation}
with $Q \in \K[x] ^{p \times p}$.
If $P$ is selected uniformly at random of degree at most $d-1$,
then $\sigma(P)$ is uniform random in $\K ^{\nu
  \times p}$. 
Additionally,
a matrix $S \in \K[x] ^{p \times p}$ is the denominator of a left coprime 
description of $B A ^{-1}$ if and only if 
$S$ is the denominator of a left coprime
description of $X (x - A_o) ^{-1}$.
\end{lemma}
\ifshortversion
\else
\ifshortversion
\vspace*{0.4cm}
\namedProof{Lemma~\ref{lem:realA} page \pageref{lem:realA}}
\else
\Proof
\fi
We first establish~(\ref{eq:real}) for $B$ the identity matrix of dimension
$n$ and for $A$ in column Popov form~\cite{Pop70} (see
also~\cite[\S6.7.2]{Kai80}):
$A$ is column-reduced, i.e. its 
leading column coefficient matrix has full rank; in each row of $A$ a unique entry has
maximum degree and is monic.
Let $d_1, d_2, \ldots, d_n$ be the column degrees of $A$, since $A$ is in
Popov form, $\nu = \sum _{i=1}^n d_i$.
We first assume that the $d_i$'s are greater than one.
We follow the lines of the realization constructions in~\cite[\S6.4]{Kai80}. 
Consider $D=\text{diag}(x ^{d_1}, x ^{d_{2}}, \ldots , x ^{d_{n}})$ and 
$\Psi = \text{diag}([1~x~\ldots x ^{d_i-1}], 1 \leq i \leq n) \in \K ^{n
  \times \nu}$.
Since $A$ is in column Popov form we have $A = D + \Psi A_L$ where 
$A_L \in \K ^{\nu \times n}$ is given by the low degree coefficients of 
the entries of $A$. 
We also define $X = \text{diag}([0, \ldots, 0, 1 ] \in \K ^{1 \times
  d_i}, 1 \leq i \leq n) 
\in \K ^{n
  \times \nu}$ and 
$D_o = \text{diag}( C_{x ^{d_1}}, C_{x
  ^{d_{2}}},
\ldots, C_{x ^{d_n}}) \in \K ^{\nu \times \nu}$ whose diagonal blocks 
are matrices companion to the diagonal entries of $D$.
One can directly check that 
$ \Psi (x - D_o) = D X$. 
Taking 
$A_o= D_o  - A_L X$ 
we get 
$\Psi (x-A_o) = \Psi (x- D_o + A_L X)  = DX + \Psi A_L X$, hence 
$ \Psi (x - A_o) =  A X$, or, in other words, 
\begin{equation}\label{eq:demrealA}
A ^{-1} \Psi = X (x -
A_o) ^{-1}.
\end{equation} 
If the row degrees of 
$P$ are strictly lower than the $d_i$'s then $P$ may be decomposed into 
$P(x)=\Psi P_o$. This leads to $A ^{-1} P = X (x - A_o) ^{-1} P_o$ and we
take
$\sigma(P)=P_o$. 
If $P$ has larger degrees, dividing $P$ by $A$  uniquely defines two
polynomial matrices $Q$ and $R$ such that 
$R=P - AQ$ and such that the row degrees of $R$ 
are less than the $d_i$'s (see \cite[Division Theorem\,6.3-15]{Kai80}).
Writing  $R = \Psi R_o$ we get  
$A ^{-1} P = A ^{-1}(AQ+R) = Q + A ^{-1}R = Q + X (x - A_o) ^{-1} R_o$ and
we take $\sigma(P)=R_o$. 

Now, if some column degrees are zero, say exactly $k$ of the $d_i$'s,
then for row and column permutations $U_l$ and $U_r$ and since $A$
is in Popov form,
we may write 
$$
U_lAU_r = \left[ \begin{array}{cc}
\bar{A} &  A_{12} \\
0 & I \end{array} \right] 
$$
where $\bar{A} \in \K[x] ^{(n-k) \times (n-k)}$  
has column degrees greater than one and $A_{12}$ is a constant matrix in
$\K ^ {(n-k) \times k}$.
Applying~(\ref{eq:demrealA}) to $\bar{A}$ we get matrices
$\bar{\Psi}$, $\bar{X}$ and $\bar{A}_o$ such that $\bar{A} ^{-1} 
\bar{\Psi} = 
\bar{X} (x - \bar{A}_o) ^{-1}$.
Hence, if $\Psi$ and $X$ are constructed by augmenting 
$\bar{\Psi}$ and $\bar{X}$ with $k$ zero rows, we get  
\begin{equation}\label{eq:demrealAb}
(A ^{-1} U_l ^{-1}) \Psi = U_r \left[ \begin{array}{cc}
\bar{A} ^{-1} &  - \bar{A} ^{-1} A_{12}\\
0 & I \end{array} \right] \Psi = U_r X (x - \bar{A}_o) ^{-1}.
\end{equation}
Then $\sigma$ may be defined as previously, 
if $R= \Psi R_o$  is the remainder of the division of 
$U_lP$ by $U_lAU_r$, then $\sigma(P)=R_o$.

If $R_o$, with $R= U_l ^{-1} \Psi R_o$, is the image of a matrix $P$ of degree
less than $d-1$, we have $P=R+AQ$. 
For another matrix $R'_0 \in \K ^{\nu \times p}$, with $R'=\Psi R'_o$, 
this defines a unique matrix $P'=R'+AQ$ of degree less than $d-1$ 
such that $\sigma(P')=R'_o$. 
Hence any two matrices  
in $\K ^{\nu \times p}$  have the same number of inverse images of
degree less that $d-1$ by
$\sigma$.
Together with the fact that the restriction of $\sigma$ to the matrices of degree less than $d-1$ 
is surjective, this shows that 
a uniform random choice of $P$ of degree less than $d-1$ 
leads to a uniform random choice $\sigma (P)$ in $\K ^{\nu \times p}$.

For general matrices $A \in \K[x] ^{n \times n}$ and 
 $B \in \K[x] ^{p \times n}$, let $V$ be unimodular such that 
$\tilde{A}=AV$ is in Popov form.
 From the above we
know that 
$$\tilde{A}(x) ^{-1} P(x) = Q(x) + X (x - A_o) ^{-1} \sigma (P).$$
Taking $\tilde{X} = B(x)V(x)X - Q_B(x) (x-A_o)$ the remainder of 
the division of $B(x)V(x)X$ by $(x-A_o)$ this leads to
$$B(x)V(x)\tilde{A}(x) ^{-1} P(x) = B(x)V(x)Q(x) + Q_B(x) +  \tilde{X} (x - A_o) ^{-1} \sigma (P)$$
which is 
$$
B(x)A(x) ^{-1} P(x) = \tilde{Q}(x) + \tilde{X} (x - A_o) ^{-1}
\sigma (P)$$
where $\tilde{Q}$ is a matrix polynomial and $X$ is a constant
matrix as the remainder of a division by a matrix of degree one.
This establishes~(\ref{eq:real}) with an appropriate $\sigma$.

It remains to show the property on denominator matrices $S$.
We use Lemma~\ref{lem:muldenom} and prove that $SB A ^{-1}$ and 
$SX (x - A_o) ^{-1}$ are polynomials for the same denominator
matrices~$S$. 
In~(\ref{eq:demrealAb}) the matrix $[I_{n-k}~0]^T$ is a submatrix of
$\Psi$. 
Therefore $\tilde{A} ^{-1}$ and $\tilde{A} ^{-1} \Psi$ are
polynomials for the same polynomial matrices $SBV$ ($B$ and $V$ are fixed), and 
$S B A ^{-1}$ and $S B A ^{-1} \Psi$ are
polynomials for the same~$S$.
Finally notice that $S \tilde{X} (x - A_o) ^{-1}$
is polynomial for the same set of matrices $S$ since 
$\tilde{X} (x - A_o) ^{-1}$ is the fractional part of $B A ^{-1} \Psi$.
\Endproof

\fi

We now state 
the required properties for~$P$, and prove them 
on the realization~(\ref{eq:real}).

\begin{proposition} \label{prop:lrHp}
Let $A \in \K[x]^{n \times n}$ be non-singular of degree less than $d$ and determinantal
degree $\nu$, and let $B \in \K[x]^{p \times n}$.
Assume that $S \in \K[x]^{p \times p}$ is any denominator of a left irreducible fraction  
description of $B A ^{-1}$. Then  
there exists a matrix $P$ of degree less than $d-1$ 
in 
$\K[x]^{n \times p}$
such that
\begin{subequations} \label{eq:denomHp}
\begin{align}
H_p = B A ^{-1} P & =   C T^{-1} \label{eq:denomHpr}\\
& =  S^{-1} N_p \in \K[x] ^{p \times p}\label{eq:denomHpl}
\end{align}
\end{subequations}
where $C T^{-1}$ is a right irreducible description with 
$T \in \K[x] ^{p \times p}$ of degree less than $\lceil \nu /p \rceil \leq (n/p)d+1$,
and where $S^{-1} N_p$ is a left irreducible description.
\end{proposition}
\Proof 
For $\nu=0$ ($BA ^{-1}$ is a polynomial), the results hold with
$T=S=I$. In the general case 
Lemma~\ref{lem:realA} gives 
$$
H_p(x) = B(x)A(x) ^{-1} P(x) = Q(x) + X (x - A_o) ^{-1} \sigma (P).
$$
For studying denominators of irreducible descriptions of $H_p$ one
can forget its polynomial part, hence 
we now focus on the fraction $X (x-A_o) ^{-1} \sigma(P)$. 
Lemma~\ref{lem:realA} also gives that there exists a left irreducible 
fraction description of $X (x - A_o) ^{-1}$ with denominator~$S$:
\begin{equation} \label{eq:leftmin}
S(x)^{-1} N'(x) = X (x - A_o) ^{-1}.
\end{equation}
Through the application $\sigma$, choosing an adequate polynomial
$P \in \K[x]^{n \times p}$ for $H_p$ reduces to choosing an adequate
constant $Y \in \K^{n \times p}$ for $X (x-A_o)
^{-1}$.

We now use the formalism of  minimum generating polynomials of matrix
sequences introduced in~\cite{Vil97-1,Vil97-2}
and~\cite[\S2]{KaVi04-2}. 
By  Lemma~2.8 in~\cite{KaVi04-2},
finding a matrix~$Y$ such that $
X (x-A_o) ^{-1} Y = C'(x) T(x) ^{-1}
$ with $T$~as expected, 
reduces to finding an appropriate~$Y$ with~$T$ the 
a right minimum generator 
of the sequence $\{ X A_o ^i Y\}_{i \geq 0}$.
From~(\ref{eq:leftmin}) we have that~$S$ is
a left minimum generator of 
$\{ X A_o \}_{i \geq 0}$. Therefore one may use the construction
of~\cite[Corollary~6.4]{Vil97-1},
together with~\cite[Theorem~2.12]{KaVi04-2}. This provides 
a $Y$ and a right minimum generator~$T$ of $\{ X A_o Y\}_{i \geq 0}$
with determinantal degree equal to the determinantal degree $\mu$ of
$S$, and  
with degree bounded by $\lceil \mu /p \rceil \leq \lceil \nu /p \rceil$.
A matrix $P$ of degree $d-1$ such that $\sigma (P)=Y$ is an
appropriate choice for concluding the proof of~(\ref{eq:denomHpr}).
Indeed, $C=QT+C'$ and $T$ gives an appropriate right irreducible
description.
The corresponding left description $S ^{-1} N_p$ is coprime
by~\cite[Lemma 6.5-6]{Kai80},
which establishes~(\ref{eq:denomHpl}). 
\Endproof 

Proposition~\ref{prop:lrHp} shows that if $P$ has symbolic entries, then 
a right coprime description of $H_p = B A ^{-1} P = C T ^{-1}$ can be found
with a denominator matrix of degree less than  $d$, and with the same
left denominators as for $B A ^{-1}$.

\begin{remark} \label{rem:randdenom}
Proposition~\ref{prop:lrHp} establishes the existence of appropriate
descriptions  $S^{-1} N_p$ and $C T ^{-1}$
for a symbolic $P$. As a consequence 
of  Lemma~\ref{lem:realA}, \cite[Corollary~6.4]{Vil97-1} or 
\cite[Section 2]{KaVi04-2}, and by
evaluation~\cite{DeMLi78,Zip79,Sch80},
the same denominator properties will hold for a random matrix $P$.
\end{remark}


\section{From compressed minimal bases to minimal bases} \label{sec:decompress}

As seen in Introduction,
we will compute a small basis for the nullspace of the input matrix
as a set 
of successive minimal bases of matrices like 
in~(\ref{eq:geneqp}).
The latter minimal bases are computed in two main steps.
We first compute the expansion of $H_p=B A ^{-1}P$ and reconstruct 
a corresponding fraction~(\ref{eq:reconintro}) with denominator $S$.
Then, if $P$ is such that $H_p$ satisfies~(\ref{eq:denomHpl}), 
we know that 
$S [BA ^{-1} ~-I_p]$ is a polynomial matrix $N$, which by construction satisfies 
$NM=0$. 

In the spirit of the scalar polynomial case and of~\cite{BeLa94} for
the matrix case, the reconstruction may be done via Pad\'e
approximation, and through the computation of particular bases  
of the nullspace of $[-I_p ~~H_p ^T]^T$.
Indeed we have the equivalence between 
$S^{-1}N_p=H_p$ 
and 
$[N_p~~S]\cdot[-I_p~~H_p ^T]^T=0$. 
Hence the purpose of this section is to identify the bases of the 
nullspace of $[H_p ^T~~-I_p]^T$ 
that actually lead to {\em minimal} bases $N$ for $M$. 

Through a
conditioning of $M$ let us first specify the location of the leading
degree terms in the latter bases (see Theorem~\ref{theo:transfer}).
 
\begin{lemma} \label{lem:Q}
For $M$ as in~(\ref{eq:geneq}) there exist a matrix $Q \in \K ^{(n+p)
  \times (n+p)}$
such that the McMillan degree of the top $n \times n$ submatrix of
$QM$ is equal to the  McMillan
degree of $QM$ (and of $M$). This implies that 
if $N$ is a minimal basis of the nullspace of 
$QM$, then $S=N_{\cdot,p+1..n+p}$ 
is row-reduced  with row degrees the Kronecker indices $\delta_1,
\ldots, \delta _p$. 
\end{lemma}
\ifshortversion
\else
\ifshortversion
\vspace*{0.4cm}
\namedProof{Lemma~\ref{lem:Q} page \pageref{lem:Q}}
\else
\Proof
\fi
If $I$ is a set of row indices such that $M_{I,\cdot}$ 
has determinantal degree $\sum
_{i=1}^{p}\delta _i$, and  let 
$Q \in \K ^{(n+p) \times (n+p)}$ be a row permutation 
$\pi$ such that $\pi(I)=\{1,\ldots,n\}$.
Then the top $n$ rows of $QM$ give the McMillan degree.
From~\cite[Theorem\,5.1\,(b)]{BLV02}
  the dominant degrees in a minimal basis
 of the nullspace of $QM$ are in the columns $\{1,\ldots,n+p\}
  \setminus \{1,\ldots,n\} = \{ n+1, \ldots , n+p\}$. 
\Endproof
\fi

\begin{remark} \label{rem:randQ}
The property given by the multiplication by $Q$  in
Lemma~\ref{lem:Q} will hold 
for a random $Q$ over $\K$ (compare to Remark~\ref{rem:randdenom}).
\end{remark}

In next sections, nullspace vectors $v^T$ for $M$ are easily obtained from
nullspace vectors $w^T$ for $QM$, indeed $v^T = w^T Q$ satisfies 
$v^TM= w^T QM=0$.
This conditioning of $M$---and implicitly of $N$---will alllow us to 
compute $S$, and then deduce $N$, from a shifted minimal basis
for the nullspace of 
$[-I_p~~H_p
^T]^T$. Shifted bases are defined as usual minimal bases by changing
the notion of degree.
For $\bar{t}$ a fixed multi-index in $\ZZ ^m$, the $\bar{t}$-degree of a vector 
$v$ in $\K[x]^m$ is 
\begin{equation} \label{eq:tdeg}
\bar{t}\text{-deg~} v = \max _{1 \leq i \leq m} \{ \deg v_i - \bar{t}_i \}. 
\end{equation}

\begin{definition}
A basis of a $\K[x]$-submodule of $\K[x]^m$, given by the rows of a
matrix $N$, is 
called $\bar{t}$-minimal if $N$ is row-reduced
with respect to the $\bar{t}$-degree. Equivalently,
$N \cdot x ^{-\bar{t}}$ is row-reduced with respect to the usual
degree
(see~\cite[Definition\,3.1]{BLV02}).
\end{definition}

For $\bar{t}=[0,\ldots,0]$ the definition corresponds to the usual definition of
minimal bases. 
The value $\bar{t}=[(d-1)_p,0_p]$  
below,
where $(d-1)_p$ and $0_p$ respectively denote the values $d-1$ and
$0$ repeated $p$ times,
 is chosen from the degree
$d-1$ of the compression matrix $P$ of Proposition~\ref{prop:lrHp}.
This value forces the row reduction in the last columns of the bases.
  
\begin{proposition} \label{prop:HpH}
Let $M \in \K[x] ^{(n+p) \times n}$ be of full rank such that 
the matrix $S$, formed by the last $p$ columns of a minimal basis $N$ 
for its nullspace, is row-reduced  with row degrees the Kronecker indices $\delta_1,
\ldots, \delta _p$.
Assume that $P \in \K[x]^{n \times p}$ 
satisfies~(\ref{eq:denomHpl}). Let $\bar{t}=[(d-1)_p,0_p] \in \NN ^{2p}$.
Then $[N_p~~S]$ is a  $\bar{t}$-minimal basis for the nullspace 
of $[-I_p~~H_p
^T]^T$ if and only if 
$N=S[B A ^{-1} ~~-I_p]=[\bar{N}~~S]$ is a minimal basis for the nullspace of $M$.
\end{proposition}
\ifshortversion
\else
\ifshortversion
\vspace*{0.4cm}
\namedProof{Proposition~\ref{prop:HpH} page \pageref{prop:HpH}}
\else
\Proof
\fi
We first prove that the condition is sufficient. If 
$[\bar{N}~~S]$ is a minimal basis for the nullspace of $M$,
the description $S^{-1}\bar{N}$ of $BA ^{-1}$ is irreducible (Lemma~\ref{lem:fracbasisred}).
The rows of $[N_p~~S]=[\bar{N}P~~S]$ are in the nullspace of  $[-I_p~~H_p
^T]^T$. 
They form a basis of the latter nullspace since otherwise 
$S ^{-1}N_p$ would not be irreducible which would
contradict~(\ref{eq:denomHpl}).
We assume that the rows of the bases are arranged by increasing degrees.
The $i$th row of $\bar{N}P$ has degree less than $\delta _i +d-1$,
hence its 
$\bar{t}$-degree is less than  $\delta _i$, which in turn is less
that the $\bar{t}$-degree
of the $i$th row of $S$.
Which shows that $[N_p~~S]$ is row-reduced with respect to
the $\bar{t}$-degree since
$S$ is row-reduced by assumption on $M$ and by
Theorem~\ref{theo:rowred}). 
The $\bar{t}$-minimality follows.

Conversely, if $[N_p~~S]$ is a $\bar{t}$-minimal basis for the nullspace
of $[-I_p~~H_p^T]^T$, 
then by~(\ref{eq:denomHpl}) 
$N=S[B A ^{-1} ~~I]=[\bar{N}~~S]$ is a polynomial matrix, and 
$N_p=\bar{N}P$. Since $[N_p~~S]$ is a basis, $S^{-1}N_p$ is
irreducible (Lemma~\ref{lem:fracbasisred}), hence also by~(\ref{eq:denomHpl}),
$S^{-1}\bar{N}$ is irreducible and  $[\bar{N}~~S]$
is a basis for the nullspace of $M$.
It remains to show that $[\bar{N}~~S]$ is row-reduced.
There exists a 
unimodular $p \times p$ matrix $U$ such that
$[\bar{N}~~S]=U[\bar{L}~~R]$
where $[\bar{L}~~R]$ is a row-reduced basis for the nullspace of
$M$, hence where $R$ is row-reduced by assymption on~$M$. By the 
predictable-degree property~\cite[Theorem\,6.3-13]{Kai80},
the degree of the $i$th row of $S$ is 
$\deg S_i = \max _{j=1,\ldots,p} \{ \delta _j + \deg U_{ij}\}$.
The degree of the $i$th row of $\bar{N}$ is 
 $\deg \bar{N}_i \leq \max _{j=1,\ldots,p} \{ \delta _j + \deg
 U_{ij}\}$.
The $\bar{t}$-degree of the $i$th row of $N_p$ may then be bounded
as follows,
$$
\bar{t}\text{-deg~} ((N_p)_i) = 
\deg ((\bar{N}P)_i) - (d-1) 
\leq \max _{j=1,\ldots,p} \{ \delta _j + \deg
 U_{ij}\} \leq \deg S_i = \bar{t}\text{-deg~} S_i.
$$
Since $[N_p ~~S]$ is row-reduced with respect to the
$\bar{t}$-degree,
this implies that $S$ itself is row-reduced. By assumption on~M the degrees of $S$ 
are dominating in $[\bar{N}~~S]$, hence the latter matrix also is row-reduced
and is a minimal basis for the nullspace of $M$.
\Endproof 
\fi

For compressing matrices $P$ which satisfy~(\ref{eq:denomHpl}),
Proposition~\ref{prop:HpH} establishes strong links
between the nullspace of $[-I_p~~H_p
^T]^T$ and the one of $M$.
In particular, $[-I_p~~H_p
^T]^T$ and $M$ have 
the same Kronecker indices. For any given $\delta$, 
there is a one-to-one correspondence 
between the vectors of $\bar{t}$-degree $\delta$ in the
nullspace of $[-I_p~~H_p
^T]^T$, and those of degree $\delta$ in the nullspace of $M$.
This is seen from the ``$S$'' common part of the bases.


\section{Computing nullspace minimal vectors}\label{sec:smalldeg}

We still consider a full column-rank matrix 
$M$ be of degree $d$ as in~(\ref{eq:geneq}). Let $\delta$ be a
fixed integer and  $\kappa (=\kappa(\delta))$ be 
the number of vectors of degree less than $\delta$ in 
a minimal basis $N$ of the $\K[x]$-nullspace of $M$.
In this section 
we study the cost for computing $\kappa$ such vectors.

\vspace*{0.4cm}
\noindent
{\bf Algorithm {\sf Nullspace minimal vectors\,($M$,$\delta$)}}\\

\noindent
\hspace*{0.5cm}\begin{tabular}{ll}
{\em Input}: & $M \in \K[x]^{(n+p)\times n}$ of degree $d$, a degree
threshold $\delta$,\\
& $M$ has full column-rank.\\
{\em Output}: & $\kappa = \max \{1 \leq i \leq p \text{~s.t.~} \delta _i
\leq \delta\}$,\\
& independent vectors $N_i \in \K[x]^{n+p}$ of degree $\delta _i$, $1 \leq i \leq \kappa$, in the
nullspace of $M$.
\end{tabular}

\vspace*{0.4cm}

\noindent
\hspace*{0.5cm}\begin{tabular}{ll}
(a)& $M:=QM$ for a random $Q \in \K ^{(n+p) \times (n+p)}$;  \\
(b) & $M:=M(x+x_0)$ for $x_0$ random in $\K$; \\
& $A:=M_{1..n,1..n}$, {\bf if} $\det A(0) =0$ then {\bf fail}; ~~~~
/*~$\text{rank}~M$ is probably less than $n$~*/\\
&  $B:=M_{n+1..n+p,1..n}$; \\
& $\eta := \delta + d + \lceil nd/p \rceil$;\\
(c) & $H:=$ expansion of $B A ^{-1} \,\bmod\, x^{\eta}$; \\
(d) & $H_p:= HP$ for $P$ random in $\K[x]^{n \times p}$ of degree
less than $d-1$; \\
& $\bar{t}= [(d-1)_p,0_p]=[d-1,\ldots d-1,0,\ldots,0] \in \NN ^{2p};$\\
(e) & $L:=[{\mathcal N}_p ~~{\mathcal S}]:=$ a $\sigma$-basis  
with respect to $\bar{t}$
for $[-I_p~~H_p^T]^T$ of order $\eta$;\\
(f)& $\kappa :=$ the number of rows of $[{\mathcal N}_p ~~{\mathcal S}]$ of
$\bar{t}$-degree at most $\delta$; \\
& select the corresponding $\kappa$ rows $S_i$ of ${\mathcal S}$ 
by increasing degrees, $1 \leq
i \leq \kappa$;\\
(g) & $N_i:=S_i [H ~~-I_p] \,\bmod\, x^{\delta +1}$, $1 \leq i \leq \kappa$; \\
& $N_i(x) := N_i(x-x_0)Q$, $1 \leq i \leq \kappa$; \\
& $\lambda := \# \{ N_i \text{~s.t.~} N_i M=0\}$\\
(h) & {\bf if} $\lambda \neq \kappa$ then {\bf fail}; ~~~~~~~~~~~
/*~certification of $\kappa$~*/\\ 
& $N^{(\delta)}:=$ the $\kappa \times (n+p)$ matrix whose rows are
the $N_i$'s;\\
(i) & {\bf if} $N^{(\delta)}$ is not row-reduced then {\bf fail}; ~~~~~~~~~
/*~certification of the minimality~*/\\
& else {\bf return} $\kappa$ and $N_i, ~1 \leq i \leq \kappa$.\Endproof\\ 
\end{tabular}

Algorithm {\sf Nullspace minimal vectors}  
starts with lifting on a compressed
matrix (Proposition~\ref{prop:lrHp}).
 Then it partially (subject to the degree threshold) 
computes a denominator matrix~$S$ through a
partial $\bar{t}$-minimal basis computation. Using 
Proposition~\ref{prop:HpH} the target nullspace vectors are finally
obtained.

We prove the algorithm and its cost in the rest of the section.
Step\,(a) is the conditioning seen in Section~\ref{sec:decompress}
to ensure the degree dominance of the last $p$ columns of $N$. 
Together with the randomized compression of Step\,(d) studied in Proposition~\ref{prop:lrHp}
this will allow the computation of $S$ at Step\,(e). 
Step\,(b) is a randomized choice for working with a matrix $A$
non-singular at $x=0$. The latter condition is required for
computing at Step\,(c) the expansion of $B A ^{-1}$ by lifting~\cite{Sto02,Sto03}.
Step\,(e) partly reconstructs a description $S ^{-1}N_p$ from a truncated 
expansion of $H_p$. The computation is explained in
Lemma~\ref{lem:approxSNp}
below, and the selection of small degree rows at Step\,(f) is
justified. Our approach for the reconstruction is very close to the column reduction 
of~\cite[\S3]{GJV03-2}.
A degree less than~$\delta$ in~$S$ corresponds to a $\bar{t}$-degree
(see~(\ref{eq:tdeg}))
less than~$\delta$ in $[N_p~~S]$ (the compression using~$P$
increases the degree in~$N_p$  by~$d-1$), and to a degree less
than~$\delta$ 
in~$N$.  
Step\,(g) applies 
Proposition~\ref{prop:HpH} for partly reconstructing the nullspace
of $M$, and Steps\,(h) and\,(i) certify the outputs.

The partial reconstruction of  $H_p$ (i.e.
the computation of a $\bar{t}$-minimal basis at
Step\,(e), and of the denominator matrix $S$ at Step\,(f))
is done using a minimal ``nullspace basis expansion''---or
$\sigma$-basis~\cite{BeLa94}.
We generalize~\cite[\S3]{GJV03-2}
and~\cite[\S4.2]{BLV02} especially for the partial computation aspects.

\begin{definition}\label{def:sigmabase}
Let $G$ be in $\K[[x]]^{q \times p}$.
Let $\bar{t}$ be a fixed multi-index in $\ZZ ^q$. 
A $\sigma$-basis of (matrix-)order $d$ with respect to $\bar{t}$
for $G$ is a matrix polynomial $L$ in
$\K[x] ^{q \times q}$ such that:
\begin{description}\vspace*{-0.15cm}
\item {\sc i})\, $L(x)G(x) \equiv 0 \,\bmod\, x^d$;
\vspace*{-0.15cm}
\item {\sc ii}) every $v \in \K[x]^q$ such that $v(x)G(x)= O(x^d)$ 
admits a unique decomposition $v ^T = \sum _{i=1}^{q} \alpha _i L_i$ 
where, for $1 \leq i \leq q$, $L_i$ is the $i$th row of $L$, and 
$\alpha _i$ is a scalar polynomial in $\K[x]$ such that 
$ \deg \alpha _i  + \bar{t}\text{\rm -deg~} L_i \leq 
\bar{t}\text{\rm -deg~} v $.
\end{description}  
\end{definition} 

The reader may notice that 
we have slightly adapted the notion of order of 
the original Definition~3.2
of~\cite{BeLa94}
for a fully matrix point of view. We also use the notion of shifted
degree
(see~\cite{BLV02}) equivalently to the notion of 
defect used in~\cite[Definition\,3.1]{BeLa94}.
The following shows that a $\sigma$-basis to sufficiently high order
contains a minimal basis.

\begin{lemma} \label{lem:nullinsigma}
Let us assume that  a minimal nullspace basis of $G$ has  $\kappa$ vectors of 
  $\bar{t}$-degree at most $\delta$, 
and consider a $\sigma$-basis $L$ with respect to $\bar{t}$.
For an approximation order greater than 
$\delta+1$, 
at least $\kappa$ rows in
  $L$ have $\bar{t}$-degree at most~$\delta$.
\end{lemma}
\ifshortversion
\else
\ifshortversion
\vspace*{0.4cm}
\namedProof{Lemma~\ref{lem:nullinsigma} page \pageref{lem:nullinsigma}}
\else
\Proof
\fi
See the proof of~\cite[Proposition\,5]{JeVi04}.
We consider the $\kappa$ rows of degree less than $\delta$ in a 
minimal nullspace basis of $G$. We order them 
by increasing degrees $\delta _1, \delta _2, \ldots , \delta _{\kappa}$.
The first row $v_1$ has degree $\delta _1$ therefore by {\sc ii})
of Definition~\ref{def:sigmabase}, $v_1$ can be written as 
$$
v_1 = \sum _{i=1}^{q} \alpha _i L_i, \text{~with~}  
\deg \alpha _i  + \bar{t}\text{\rm -deg~} L_i \leq 
\bar{t}\text{\rm -deg~} v_1 = \delta _1.
$$
We deduce that one row of $L$ has $\bar{t}$-degree at most
$\delta_1$. 
Now if $L$ has~$i-1$ rows of degrees $\delta _1, \ldots , \delta _{i-1}$,
with $v_{i}$ of $\bar{t}$-degree $\delta _{i}$, then  
the same reasoning as for $v_1$ shows that $L$ has a row of degree 
less than $\delta _{i}$, linearly independent with respect to the first~$i-1$
chosen ones. 
The proof is concluded with $i=\kappa$.
\Endproof
\fi

Next lemma identify the situation when a $\sigma$-basis will give the 
exact information we need. We assume that we are in the situation of
Proposition~\ref{prop:HpH}, in particular $S$ in the minimal bases 
has row degrees $\delta _1, \ldots , \delta _p$, the Kronecker
indices of~$M$ and of $[-I_p ~~H_p^T]^T$.
We fix a value $\delta$ 
and define $\kappa = \max \{1 \leq i \leq p \text{~s.t.~} \delta _i
\leq \delta\}$, and $\bar{t}=[(d-1)_p,0_p] \in \NN ^{2p}$.

\begin{lemma} \label{lem:approxSNp}
Let us assume we are in the situation of
Proposition~\ref{prop:HpH}.
Let $L$ be a $\sigma$-basis for $[-I_p ~~H_p^T]^T$, with respect to 
$\bar{t}$, 
and of order of approximation at least $\eta = \delta + d + \lceil nd/p \rceil$.
Then exactly $\kappa$ rows of $L$ have $\bar{t}$-degree at
most~$\delta$,
are in the nullspace of $[-I_p ~~H_p^T]^T$, and  
have $\bar{t}$-degrees $\delta _1, \ldots , \delta
_{\kappa}$. 
\end{lemma}
\ifshortversion
\else
\ifshortversion
\vspace*{0.4cm}
\namedProof{Lemma~\ref{lem:approxSNp} page \pageref{lem:approxSNp}}
\else
\Proof
\fi
We generalize the proof of~\cite[Lemma\,3.7]{GJV03-2} 
to the partial computation of the basis, and to the shifted case.
We first verify that at most $\kappa$ rows of 
$L$ have $\bar{t}$-degree less than $\delta$, then we  prove
their existence. Note that the rows of $L$ are linearly independent~\cite{BeLa94}.

Let $L_i =[\bar{L}_i ~~ S_i]
\in \K[x]^{2p}$ be a row of $L$ of $\bar{t}$-degree at
most~$\delta$.
From {\sc i}) in Definition~\ref{def:sigmabase},
$$
S_i(x) H_p(x) \equiv \bar{L}_i(x)  \,\bmod\, x^{\eta},
$$
and from the assumption~(\ref{eq:denomHpr}) on $P$,
\begin{equation} \label{eq:fracbothsides}
S_i(x) C(x) \equiv \bar{L}_i(x)T(x)  \,\bmod\, x^{\eta}.
\end{equation} 
We now look at the degrees in both sides of latter identity.
We have $\deg S_i = \bar{t}\text{-deg}\, S_i \leq \delta$.
By assumption on $M$, the degree of $B A ^{-1}$ is at most zero,
hence the degree of $H_p$ is at most $d-1$. The latter is also true  for 
$C T^{-1}$ in~(\ref{eq:denomHpr}), which implies that $\deg C \leq
\deg T + d-1$. Using the degree bound on $T$ in
Proposition~\ref{prop:lrHp},
the left side term of~(\ref{eq:fracbothsides}) thus
have degree at most $\eta-1$. In addition,  
$\deg \bar{L}_i = \bar{t}\text{-deg}\, (\bar{L}_i) + (d-1)  \leq
\delta + d-1$. Hence both sides in~(\ref{eq:fracbothsides}) have
degree at most $\eta -1$ and we deduce that 
\begin{equation} \label{eq:eqbothsides}
S_i(x) C(x) = \bar{L}_i(x)T(x).
\end{equation} 
It follows that $L_i =[\bar{L}_i ~~ S_i]$ is in 
nullspace of $[-I_p ~~ H_p^T]^T$. 
Using the equivalence with
the nullspace of~$M$ in Proposition~\ref{prop:HpH}, 
one may associate to $L_i = [\bar{L}_i ~~ S_i]$ a row vector 
$N_i = [\bar{N}_i ~~ S_i]$, with $\bar{N}_i P = \bar{L}_i$,
of degree less than
$\delta$ in the nullspace of
$M$ (the ``S'' part is row-degree dominant). Since the rows
$L_i$ are linearly independent, the  rows $N_i$ of degree less
than $\delta$, corresponding 
to the $L_i$'s of $\bar{t}$-degree 
less than $\delta$, are linearly independent. 
At most $\kappa$ such rows can exist.

We now show that $\kappa$ rows of $\bar{t}$-degree at most $\delta$
exist in $L$. We consider the $\kappa$ rows of degrees $\delta _1,
\ldots,
\delta _{\kappa}$ in a minimal basis $N=[\bar{N}~~S]$ of the nullspace
of $M$. 
They give $\kappa$ rows of $\bar{t}$-degree at most $\delta$
in the nullspace of $[-I_p~~H_p^T]^T$. 
Using Lemma~\ref{lem:nullinsigma} they lead to 
 $\kappa$ rows of $\bar{t}$-degree at most $\delta$ in $L$,
which 
are in the nullspace by~(\ref{eq:eqbothsides}),
hence their $\bar{t}$-degrees are $\delta_1, \ldots , \delta
_{\kappa}$ by minimality.
Note that the linear independency in $N$ is preserved for the 
nullspace of $[-I_p~~H_p^T]^T$
since the column-reduced part $S$ is in common. 
\Endproof

\fi

\begin{proposition} \label{prop:pnullspace}
Let $M \in \K[x]^{(n+p)\times n}$ be of full column-rank with 
Kronecker indices $\delta _1, \ldots , \delta _p$.
Algorithm {\sf Nullspace minimal vectors} with inputs $M$ and 
$\delta \in \NN$ returns $\kappa= \max \{1 \leq i \leq p \text{~s.t.~} \delta _i
\leq \delta\}$, and $\kappa$ first minimal vectors of the
nullspace of $M$.
The algorithm is randomized, it either fails or returns correct
values (Las Vegas
fashion).
\end{proposition}
\Proof
We first verify that if the random choices of $x_0$, $Q$ and $P$
work as expected then the result is correct.
We will then prove that if the algorithm does not return fail
then we are in the previous case. 
Note that the random shift $x_0$ does not modify the problem.
Indeed, $\text{rank\,}M(x)=\text{rank\,}M(x+x_0)$, and since a matrix 
whose rows form a  
minimal basis is irreducible, the Kronecker indices are invariant
under a shift.

Using Lemma~\ref{lem:Q}, the role of $Q$ is twofold: the top
$n \times n$ submatrix of $M$ becomes non-singular,
and the dominant degrees in the nullspace are in the last columns. 
If $\det A(0) \neq 0$ then the rest of the computation is valid,
in particular the expansion of $B A ^{-1}$ at Step\,(c).
The basis $L$ of order~$\eta$ as required can be computed from the expansion 
of $B A ^{-1}$ to the order~$\eta$~\cite{BeLa94,GJV03-2}.
If the choices of $Q$ and $P$ are successful then  
Lemma~\ref{lem:approxSNp}
ensures that the value of $\kappa$ is correct; the corresponding
rows of $L$ are in the nullspace of $[-I_p ~~ H_p^T]^T$.
The nullspace correspondence of Proposition~\ref{prop:HpH}
then shows that $S_i [ BA ^{-1}~~-I_p]$ is a polynomial row of degree
less than $\delta$, hence the computation of $N_i$ can be done
modulo $x ^{\delta +1}$.

We now study the certification of the outputs. If $\det A(0) \neq 0$ then
we know that $M$ has full column-rank.
The algorithm may then potentially fail with respect to the output value
$\kappa$, there could actually be less or more minimal vectors of degrees at
most $\delta$. It may also fail with respect to the minimality of
the output vectors. In any case, the computation of $\lambda$
ensures that the returned $N_i$'s are in the nullspace. 

To avoid confusion we now denote by $\kappa _o$ the output value and
keep $\kappa$ for the correct (unknown) value. Let us first see that
$\kappa _o \geq \kappa$. Indeed, to the $\kappa$ rows 
$N_i = [\bar{N}_i~~S_i]$ of degree less than $\delta$ 
is in the nullspace of $M$, one may associate $\kappa$ rows
$[\bar{N}_iP~~S_i]$
of $\bar{t}$-degree less than $\delta$ in the nullspace of 
$[-I_p ~~ H_p^T]^T$.
Since $\det A \neq 0$, we know that the $S_i$'s are linearly
independent.
Hence we have $\kappa$ linearly independent rows
of $\bar{t}$-degree less than $\delta$ in the nullspace of 
$[-I_p ~~ H_p^T]^T$.
Then by Lemma~\ref{lem:nullinsigma}, 
there must be $\kappa _o \geq \kappa$ rows of $\bar{t}$-degree less
than $\delta$
in~$L$.
If $\lambda = \kappa _o$ then we have found $\kappa_o$ linearly
independent rows (from the $S_i$'s) of degree less than $\delta$ 
(the degree is forced by construction at Step\,(g)) 
in the nullspace of
$M$ (test at Step\,(h)), hence $\kappa _o > \kappa$ cannot happen, and $\kappa _o =
\kappa$. The returned value $\kappa$ is always correct.
In the latter case the 
returned vectors are linearly independent in the nullspace and satisfy the degree  
constraint.

We finally show that the returned vectors must be minimal.
The corresponding $\kappa$ rows, say $L_i = [\bar{L}_i ~~S_i]$ 
for $1 \leq i \leq \kappa$, in the nullspace
of $[-I_p~~H_p^T]^T$, must be minimal. Otherwise, 
by {\sc ii}) of Definition~\ref{def:sigmabase}, a row of smaller
$\bar{t}$-degree would have been selected in $L$. In particular,
the matrix formed by the $\bar{L}_i$'s and the one formed by the
$S_i$'s
 are left relatively prime (no common left divisor other than
 unimodular). 
The $\kappa$
computed rows $N_i = [\bar{N}_i ~~ S_i]$ satisfy
$\bar{N_i}P=\bar{L}_i$, hence the matrix formed by the $\bar{N}_i$'s
and 
the one formed by the $S_i$'s
are also left relatively prime.
Let $N_o^{(\delta)}$ be the $\kappa \times (n+p)$ matrix 
whose rows are the computed $N_i$'s, and let 
 $N^{(\delta)}$ be a $\kappa \times (n+p)$
matrix whose rows are $\kappa$ first minimal vectors for the
nullspace of $M$. Then the primality implies that 
there exist a unimodular $U$ such that $N_o^{(\delta)} = U
N^{(\delta)}$ (see Remark~\ref{rem:submodule}).
Therefore the rows of $N_o^{(\delta)}$ have minimal degrees if and
only if $N_o^{(\delta)}$ is row-reduced.
The check is made at Step\,(i).
\Endproof

From the arguments used in the proof of
Proposition~\ref{prop:pnullspace} we see that the algorithm may
fail because  
the computed value $\kappa$ is too large. 
This will essentially happen  for bad choices of $P$, when the
nullspace of the compressed matrix (see (\ref{eq:denomHpl})), and the approximating 
$\sigma$-basis (see (\ref{eq:denomHpr})), 
does not reflect the nullspace of $M$ correctly.
Then, even for correct values of $\kappa$, the minimality may not be 
ensured
without the test at Step\,(i).
A bad choice of $Q$, depending on $P$, may lead to a row reduction
in the non-dominant part of the basis (see
Lemma~\ref{lem:Q}), 
and to a loss of minimality
(see Proposition~\ref{prop:HpH})\footnote{An improvement would be to
combine both conditionings into a unique one with three different
effects,
left and right fractions for $H_p$, and location of the dominant degrees.}.
A correctly computed value of
$\kappa$ may lead to a smaller value $\lambda$ after the
truncation\,(g)
of a non-minimal vector.

We also note that the minimality condition could be relaxed in the
algorithm. Avoiding the last certificate would lead to the Las Vegas
computation of $\kappa$ independent vectors (possibly non-minimal) in the nullspace. 

\begin{corollary} \label{cor:costpnullspace}
Let $M \in \K[x]^{(n+p)\times n}$ be of full column-rank and degree $d$
with $1 \leq p \leq 2n$, 
and let $d \leq \delta \leq nd$.
Minimal independent vectors in the nullspace of $M$,
of degrees the Kronecker indices less than $\delta$, can be computed 
by a randomized Las Vegas (certified) algorithm in 
$$
O(\lceil {p\delta}/{nd}\rceil\,\MM(n,d) \log n
+(n/p)  \MM(p,\delta)
+ \MM(p,\delta) \log \delta
+ n^2 \Pgcd(d) \log n) 
$$
operations in $\K$.
The cost 
is 
\begin{equation}\label{eq:costcompress}
O( \MM(n,d) \log (nd) + n^2 \Pgcd(d) \log n + n \PM(nd))
\end{equation} 
when $p\delta /(nd) =O(1)$.
\end{corollary}
\ifshortversion
\else
\ifshortversion
\vspace*{0.4cm}
\namedProof{Corollary~\ref{cor:costpnullspace} page \pageref{cor:costpnullspace}}
\else
\Proof
\fi
We use either~(\ref{eq:costMM1}) or~(\ref{eq:costMM2}), and the
corresponding simplifications~(\ref{eq:MMrec}) for studying the cost
of Algorithm {\sf Nullspace minimal vectors}.

Steps\,(a) and\,(b) uses $O(\MM(n,d) + n^2 \PM(d))$
  operations.
From~\cite[Proposition\,15]{Sto03}, the cost for computing the
expansion $H$ is $O(\log (\eta /d) \lceil p \eta /nd \rceil \MM(n,d)
+\MMrecas(n,d))$. 
This gives 
$O(( \lceil ({p\delta})/({nd})\rceil\,\MM(n,d)  +
n^2 \Pgcd(d)) \log n)$ for $\eta = O(\delta + nd/p)$.
Step\,(d) is a polynomial matrix multiplication that can be done in 
$O(\MM(n,d))$ operations.
For the computation of the $\sigma$-basis at Step\,(e) we use the 
algorithm of~\cite[\S2]{GJV03-2} based on 
polynomial matrix multiplication. The corresponding cost from~\cite[Theorem\,2.4]{GJV03-2}
is 
$O(\MMrecgjv(p,\eta)+ \eta \MM(p))$, hence $O(\MM(p,\delta) \log
\delta)$,  or  $O(\MM(n,d) \log
d)$ if  $p\delta /nd =O(1)$.
Step\,(g) is a polynomial matrix multiplication
modulo~$x^{\delta+1}$ that can be computed in 
$(n/p)\MM(p,\delta)$ operations, this is 
$O(\MM(n,d) \log d + n \PM(nd))$ when $p\delta /nd =O(1)$.
The shift of the $N_i$'s is done in at most $O(\sum _{i=1}^p n
\PM(\delta_i))$ operations, which is less than $O((n/p) \MM(p,\delta))$, or than
  $O(n \PM(nd))$ for  $p\delta /nd =O(1)$.
The subsequent multiplication by $Q$ has lower cost.
Then we compute $N_iM$ for $1 \leq i \leq \kappa$, where $N_i$
has degree $\delta _i$, and $\sum _{i=1}^p \delta _i \leq nd$. 
Doing this computation directly as the product of a $\kappa \times
(n+p)$ matrix with possibly large degrees, by an $(n+p) \times n$ matrix of
degree $d$ would be too expensive. Instead,  
we split the large degree entries 
of the $N_i$'s and form an $n \times (n+p)$ matrix $\tilde{N}$ of degree
$d$, and recover the products $N_iM$ from the multiplication
$\tilde{N}M$.
The corresponding cost is $O(\MM(n,d))$. The final check\,(i) is done
in $O(n^{\omega}+n^2d)$ operations.
\Endproof
\fi

We see from~(\ref{eq:costcompress}) that computing vectors in the
nullspace at essentially the cost of multiplying two polynomial
matrices
relies on the {\em compromise between~$p$ and~$\delta$}. 
The algorithm is a combination of {\em matrix lifting} and {\em matrix fraction reconstruction}.
Many vectors of
small degrees are computed using lifting to a limited order and
large matrix reconstruction.
Conversely, few vectors of large degrees are computed from a
high-order 
lifting and reconstruction 
with matrices of small dimensions.

\begin{remark}\label{rem:bigp} 
Note that the random compression $P$ is introduced for $p < n$.
Still, the algorithm is proven for $p \geq n$. In the latter case however,
for the sake of simplicity, one may work directly with $H_p=H$ at Step\,(d). 
\end{remark}


\section{Small degree nullspace basis computation} \label{sec:wholebasis}

Corollary\,\ref{cor:costpnullspace}
which uses for~(\ref{eq:costcompress}) a compromise between $p$ and $\delta$, 
does not directly allow a low-cost computation of large degree vectors in a 
nullspace of large dimension.
For the latter situation, 
and for computing a whole set of linearly independent vectors in the
nullspace of a matrix $M$ in $\K[x]^{(n+q) \times n}$,  
we need to successively restrict ourselves to 
smaller nullspace dimensions (while increasing the degree).
Here we take the notation $m=n+q$ for $M$ as in~(\ref{eq:geneq}).
We keep
the notation $p$ for submatrices~(\ref{eq:geneqp}), and
successive compressions, as in Sections~\ref{sec:MFD}-\ref{sec:smalldeg} .


\subsection{Full column-rank and $n < m \leq 2n$ case} \label{subsec:2n}

Let $M \in \K[x] ^{(n+q) \times n}$ with $1 \leq q \leq n$ be of degree $d$ and
rank $n$.
The way we restrict ourselves to smaller nullspaces is derived 
from the following observation.
Let $C$
be in $\K ^{(n+p) \times (n+q)}$ with $1 \leq p \leq q$.
If $CM \in \K[x] ^{(n+p) \times n}$ also has full column-rank, then let 
$\delta _1, \ldots , \delta _p$ be its Kronecker indices, and with
the degree threshold 
$\delta = 2nd/p$ take
$\kappa (\delta) = \max \{1 \leq i \leq p \text{~s.t.~} \delta _i
\leq \delta\}$.
Since $\sum _{1}^p \delta _i \leq
nd$,
at most $nd/\delta =p/2$, hence $\lfloor p/2 \rfloor$, vectors in a minimal basis of the nullspace
of $CM$ may have degrees more than 
$\delta$, therefore $\kappa(\delta) \geq \lceil p/2 \rceil$.
From at least $p/2$ minimal vectors $D_1, \ldots ,
D_{\kappa} \in \K[x] ^{n+p}$  of degrees at most~$2nd/p$
in the nullspace of $CM$, we obtain  
$\kappa$ corresponding vectors  
$N_i = D_iC \in \K[x]^{n+q}$ in the nullspace of $M$.

\vspace*{0.4cm}
\noindent
{\bf Algorithm {\sf Nullspace}$_{2\text{\sf n}}$($M$)}\\

\noindent
\hspace*{0.5cm}\begin{tabular}{ll}
{\em Input}: & $M \in \K[x]^{(n+q)\times n}$ of degree $d$, \\
& $M$ has full column-rank and $1 \leq q \leq n$.\\
{\em Output}: & $q$ ``small'' linearly independent polynomial vectors in the nullspace of
$M$. 
\end{tabular}

\vspace*{0.4cm}

\noindent
\hspace*{0.5cm}\begin{tabular}{l}
$M:=QM$ for a random $Q \in \K^{(n+q) \times (n+q)}$;\\
{\bf if} $\det M_{1..n,1..n}(x_0) =0$ for $x_0$ random in $\K$ then {\bf fail};\\
$I = \{ \}$;\\
$p:=q$;\\
{\bf while} $\# I < q$ 
\end{tabular}\\
\hspace*{0.5cm}\begin{tabular}{ll}
~~~(a)& $\{i_1, \ldots , i_p\}:= \{n+1,
\ldots ,n+q\} \setminus I$;\\
~~~(b)& $\delta := 2nd/p$;\\
~~~(c)& construct $C \in \K ^{(n+p) \times (n+q)}$ with $C_{i,i}:=1$, $1 \leq i
\leq n$, $C_{n+j,i_j}:=1$, $1 \leq j
\leq p$, \\
&~~~~and $C_{i,j}:=0$ otherwise;\\
~~~(d)& $\bar{M}:=CM \in \K[x] ^{(n+p) \times n}$;\\ 
~~~(e)& $\{\kappa, \{D_i, 1 \leq i \leq \kappa\}\}:=$ {\sf Nullspace
  minimal vectors}\,($\bar{M},\delta$);\\
& $N^{(\delta)}_i=D_iC$, $1 \leq i \leq \kappa$;\\
~~~(f)& $N^{(\delta)}:=$ the $\kappa \times (n+q)$ matrix whose rows are
the $N^{(\delta)}_i$'s;\\
~~~(g)& $J:= \kappa$ column indices greater than $n+1$ such that 
$N_{1..\kappa,J}$ is non-singular;\\
~~~(h)& $I:= I \cup J$, $p:=p-\kappa$;\\
~~~(i)& $N:= [N^T~~(N^{(\delta)})^T]^T$; ~~~~/*~update the
nullspace~*/~~
\end{tabular}\\
\hspace*{0.5cm}\begin{tabular}{l}
$N:=NQ$;\\
{\bf return} $N_i$, $1 \leq i \leq q$.
~~~~~~~~~~~~~~~~~~~~~~~~~~~~~~~~~~~~~~~~~~~~~~~~~~~~~~~~~~\Endproof\\ 
\end{tabular}

Algorithm {\sf Nullspace}$_{2\text{\sf n}}$ is proven in
Proposition~\ref{prop:nullspace} below. Let us first give the
general idea. 
For computing the whole nullspace, the algorithm generates a sequence of decreasing 
dimensions $p$ at Step\,(h).
Using the observation made previously, each time the algorithm
passes through the ``while loop'' the dimension is divided by at
least two, hence at most $O(\log _2 q)$ stages are necessary.
This corresponds to $O(\log _2 q)$ calls to {\sf Nullspace minimal
  vectors} with input $CM$.
Each time the dimension is decreased, the degree threshold is
increased in the same proportion at Step\,(b), we preserve the invariant 
\begin{equation} \label{eq:pdeltand}
p\delta / (nd) =  2.
\end{equation}
The latter identity  will be used for applying the cost
estimate~(\ref{eq:costcompress})
of Corollary~\ref{cor:costpnullspace}.

The proof of Proposition~\ref{prop:nullspace} will check that 
$q$ vectors in the nullspace are actually computed. In addition, 
the algorithm has to ensure their linear independency.
The latter is done on the fly, and will first rely on the 
initial conditioning with $Q$ for working with a 
top $n \times n$ non-singular submatrix.
The vectors for updating the nullspace 
are computed at Step\,(e) and Step\,(f)  in 
the nullspace of $M_{\bar{I},1..n}$, with 
$\bar{I} = \{ 1, 2, \ldots, n, i_1, i_2, \ldots,
i_p\}$. This is done through the construction of the compression 
matrix $C$ at Step\,(c) which selects the corresponding rows of
$M$. The choice of the indices  $\{ i_1, i_2, \ldots,
i_p\}$ at Step\,(a), complements the index choices
at Step\,(g) that are kept in $I$ at Step\,(h) for previous stages, and will provide the 
linear independency by construction. Another perhaps simpler
 strategy for ensuring
independency could be based on randomization.

Our approach is ``greedy'', all vectors of degree under the
threshold $\delta$ are kept. It is unclear how using a formal
``divide and conquer'' would make a difference.

\begin{proposition} \label{prop:nullspace}
Let $M \in \K[x]^{(n+q)\times n}$ with $1 \leq q \leq n$ be of full
column-rank.
Algorithm {\sf Nullspace}$_{2\text{\sf n}}$
computes $q$ linearly independent polynomial vectors in the 
nullspace of $M$.
If $M$ has degree $d$ then the sum of the degrees of the output
vectors is less than  $nd\lceil \log _2 q
\rceil$.
The algorithm is randomized, it either fails or returns correct
values (Las Vegas
fashion).
\end{proposition}
\Proof
The initial multiplication by $Q$ and the corresponding failure test
ensure that the top $n \times n$ matrix of $M$ is invertible 
when the algorithm enters the ``while loop'' (if the algorithms
fails then $M$ probably has rank less than~$n$).
At Step\,(f), $\kappa$ vectors in the nullspace of $M$ are computed,
indeed, $D_i\bar{M} = D_iCM=0$ directly gives 
$N^{(\delta)}_iM=D_iCM=0$. 
The number of elements of $I$ is increased by $\kappa$ at Step\,(h), hence
is equal to the current total number of computed vectors. 
Since $\kappa \leq q - \#I$, if the algorithm terminates then 
exactly $q$ nullspace vectors are obtained.
In addition we have already seen that $\kappa$ is at least $\lceil p/2 \rceil$,
therefore, if we denote by $p_{\text{new}}$ the new value of $p$  at
Step\,(h), we 
have $p_{\text{new}} \leq \lfloor p/2 \rfloor$, which means that the
algorithm terminates after having passed  
through the ``while loop'' at most $\lceil \log _2 q \rceil$ 
  times.

Algorithm {\sf Nullspace minimal vectors} returns $\kappa$ linearly
independent row vectors $D_i \in \K[x]^{n +p}$ at Step\,(e).
Let $D$ be the $\kappa \times (n+p)$ matrix whose rows are the
$D_i$'s.
We respectively denote the $k$th column of $N  ^{(\delta)}$ and $D$, by 
$N_{\cdot,k}  ^{(\delta)}$ and $D _{\cdot,k}$. The constrution of $C$
leads to:
\begin{subequations} \label{eq:colND}
\begin{align}
N _{\cdot,i_j}  ^{(\delta)} &=  D _{\cdot, n+j}, \text{~if~} 1 \leq j
\leq p,\label{eq:colND1}\\
N _{\cdot,k}  ^{(\delta)} &= 0, \text{~otherwise}.\label{eq:colND2}
\end{align}
\end{subequations}
Since the top $n \times n$ matrix of $M$, and consequently the one
of $\bar{M}$, is non-singular, $\kappa$ linearly independent
columns may be found among the last $p$ columns of $D$.
Therefore, from~(\ref{eq:colND1}), $\kappa$ linearly independent
columns $J$ may be found among the columns $i_1, \ldots , i_p$ of $N^{(\delta)}$.
This shows that Step\,(g) is valid.
In addition,  at subsequent stages, from Step\,(a) and~(\ref{eq:colND2}), the non-zero columns 
involved  between $n+1$ and $q$ will be outside $J$, the
corresponding nullspace vectors will thus be
linearly independent from $N_i ^{(\delta)}$, $1 \leq i \leq \kappa$. 
At each stage the $N_i ^{(\delta)}$'s are linearly independent,
and are independent from those computed subsequently,
hence we have proven that the algorithm returns $q$ linearly
independent nullspace vectors.

Each of the times the algorithm  passes through the ``while loop'',
the sum of the degrees of the computed vectors is bounded by the sum
$nd$ of the Kronecker indices. Indeed, these vectors are minimal for
the nullspace of the submatrix $\bar{M}$.
Hence the sum of the degrees in output is  
less than $nd\lceil \log _2 q
\rceil$.
\Endproof

The computed vectors $D_i$'s are minimal in the nullspace of $CM$
but the minimality is not preserved in general for the vectors $N_i$'s in
the nullspace of $M$. 
The output basis for the nullspace as $\K (x)$-vector
space may not be a basis for the $\K [x]$-module. 
However,  Proposition~\ref{prop:nullspace}
shows that if the sum of the Kronecker indices is $nd$ (the maximum
possible), then the sum of the computed degrees is only within
$\lceil \log _2 q \rceil$ times the optimum.
We notice also that the vectors computed at the first stage are
minimal vectors by Proposition~\ref{prop:pnullspace}, hence the
algorithm reaches the optimum for a generic matrix $M$ (the whole
nullspace is computed with $p=q$).
It would be interesting to study the loss of minimality compared to
the Kronecker 
indices in the general case. 

We also remark that the algorithm could be slightly modified 
for computing a row-reduced nullspace matrix $N$.
The intermediate bases matrices $D\in \K[x]^{\kappa\times (n+p)}$ 
whose rows are the $D_i$'s are row-reduced by
Proposition~\ref{prop:pnullspace}.
By Lemma~\ref{lem:Q} the dominant degrees are in the last~$p$
columns. The column index selection of Step\,(g) may be specialized
for choosing indices corresponding to dominant degrees. From there, the
proof of Proposition~\ref{prop:nullspace} for establishing that the
computed vectors are independent may be extended to the fact that
the output matrix $N$ is row-reduced. This could be certified at the
end of the Algorithm {\sf Nullspace}$_{2\text{\sf n}}$ as done at Step\,(i) of
Algorithm
{\sf Nullspace minimal vectors}.

\begin{corollary} \label{cor:costnullspace2n}
Let $M \in \K[x]^{(n+q)\times n}$ be of full column-rank and degree $d$
with $1 \leq q \leq n$, $q$ polynomial vectors whose degree sum is
less than  $nd\lceil \log _2 q
\rceil$ can be computed in 
\begin{equation}\label{eq:cost2n}
O( (\MM(n,d) \log (nd) + n^2 \Pgcd(d) \log n + n \PM(nd)) \log q )
\end{equation} 
operations in $\K$ by a 
randomized Las Vegas (certified) algorithm.
\end{corollary}
\ifshortversion
\else
\ifshortversion
\vspace*{0.4cm}
\namedProof{Corollary~\ref{cor:costnullspace2n} page \pageref{cor:costnullspace2n}}
\else
\Proof
\fi
We study the cost of Algorithm {\sf Nullspace}$_{2\text{\sf n}}$.
The conditioning with the matrix $Q$ and the failure test use at most 
$O(\MM(n,d))$ operations.
We claim that the dominating cost is the body of the loop is 
the call to Algorithm {\sf Nullspace minimal vectors}. Since $O(\log
q)$ calls are sufficient, and since $p\delta/(nd)=2$,
 (\ref{eq:cost2n}) is a consequence of~(\ref{eq:costcompress})
in Corollary~\ref{cor:costpnullspace}.
Step\,(d) is the extraction of a submatrix.  The computations $N_i
^{(\delta)}=D_i C \in \K[x] ^{n+q}$, for $1 \leq i \leq \kappa$, can be done 
in $O(n^2d)$ since the degree sum of the $N_i
^{(\delta)}$'s is less than $nd$.   
 The choice of $\kappa$ column indices at Step\,(g) can be made in
 $O(n^{\omega} +n^2d)$ operations. 
\Endproof

\fi


\subsection{General case} 

We now work with a general matrix $M \in \K[x]^{m \times n}$ of
degree~$d$. 
We compute the rank $r$ of $M$ and $m-r$ linearly independent
and ``small'' polynomial vectors in the nullspace.
Our strategy first uses Monte Carlo techniques for computing
a value $\ro \leq r$, equal to $r$ with high probability.

\begin{lemma} \label{lem:Mnr}
Let $M$ be in $\K[x]^{m \times n}$ of degree $d$. 
A matrix $\tilde{M} \in \K[x]^{m \times \ro}$
  of degree $d$ and full column-rank with $\ro \leq r$, such that with high probability $\ro=r$ and 
its nullspace
is equal to the nullspace of $M$, 
can be computed in 
$O(nm\MM(r,d)/r^2)$ operations in $\K$ by a randomized Monte
Carlo (non-certified) algorithm.
\end{lemma}
\ifshortversion
\else
\ifshortversion
\vspace*{0.4cm}
\namedProof{Lemma~\ref{lem:Mnr} page \pageref{lem:Mnr}}
\else
\Proof
\fi
The matrix $M$ can be evaluated at a random value $x_0$ in $\K$ 
in $O(mnd)$ operations. With high probability the rank is preserved.
Then the rank $\ro \leq r$ after evaluation 
can be computed over $\K$ in  $O(nmr^{\omega-2})$
operations (see~\cite{KeGe85} and~\cite[Chapter\,3]{Sto00}).
We compute $\tilde{M}=MR$ for $R$ a random $n \times \ro$ matrix 
over $\K$ in $O(nm\MM(r,d)/r^2)$. 
\Endproof
\fi


Lemma~\ref{lem:Mnr} reduces the problem to the full
column-rank case.
We then apply the results of previous sections 
for computing $m-\ro$ candidate independent vectors in the nullspace
of $\tilde{M}$.
We finally test by multiplication whether the $m-\ro$ vectors are
actually in the nullspace of $M$. A positive answer implies that $r \leq
\ro$, therefore certifies that $r=\ro$, and that a correct
nullspace representation has been constructed.

The case $m \leq 2\ro$ 
has been treated in Section~\ref{subsec:2n}. It remains to handle in
particular 
the situation 
$m \gg \ro$. The sum of the Kronecker indices is at most 
$\ro d$, hence at most $\ro$ vectors may have
degrees greater than $d$.
For $m > 2\ro$,
we apply the technique of successive row indices selection of
Section~\ref{subsec:2n}
for computing $m -2\ro$ independent vectors of degrees less than $d$,
and will terminate by computing $\ro$ vectors of possibly higher
degrees using the case $m =2\ro$.

\vspace*{0.4cm}
\noindent
{\bf Algorithm {\sf Nullspace}($M$)}\\

\noindent
\hspace*{0.5cm}\begin{tabular}{ll}
{\em Input}: & $M \in \K[x]^{m\times n}$ of degree $d$. \\
{\em Output}: & $r=\text{rank}\,M$,\\
& $m-r$ ``small'' linearly polynomial vectors in the nullspace of
$M$. 
\end{tabular}

\vspace*{0.4cm}

\noindent
\hspace*{0.5cm}\begin{tabular}{ll}
(a) & compute $\ro$ and $\tilde{M}=MR \in \K[x]^{m \times \ro}$ using Lemma~\ref{lem:Mnr}; \\ 
& {\bf if} $m=\ro$ then {\bf return} $m$ and $\{\}$;\\
& $q:=\lceil (m-2\ro)/\ro \rceil$;\\
(b)& randomly ensure that the top $\ro \times \ro$ submatrix of $\tilde{M}$ is
non-singular or {\bf fail};\\
(c)& $\{ N_i, 1 \leq i \leq m-2\ro\}:=$  {\sf Nullspace minimal
  vectors}($\tilde{M}^{(k)},d$), $1 \leq k \leq q$;\\
(d)& 
$\{ N'_i, 1 \leq i \leq \min\{m,2\ro\}-\ro\}:=$
{\sf
  Nullspace}$_{2\text{\sf n}}$($\tilde{M}^{(q+1)}$);\\
& $N$ in $\K[x] ^{(m-\ro) \times m}$ the matrix whose rows are the $N_i$'s
and the $N'_i$'s;\\
(e)&{\bf if} $NM \neq 0$ then {\bf fail};\\
& else {\bf return} $\ro$ and $N_i$, $1 \leq i \leq m-\ro$.
\Endproof\\ 
\end{tabular}

For the first $m -2\ro$ vectors of degrees less than $d$ we work in 
$q = \lceil (m-2\ro)/\ro \rceil$
stages, and successively consider submatrices 
$\tilde{M}^{(1)}, \ldots , \tilde{M}^{(q)} \in \K[x] ^{\iota \times \ro}$
of $\tilde{M}$, with $2\ro < \iota \leq 3\ro$.
More precisely, $\tilde{M}^{(k)} \in \K[x] ^{3\ro \times \ro}$ for 
$1 \leq k \leq q-1$, and $\tilde{M}^{(q)} \in \K[x] ^{(m-(q-1)\ro) \times \ro}$.
Like in Algorithm {\sf Nullspace}$_{2\text{\sf n}}$
we always ensure by randomization that 
the top $\ro \times \ro$ submatrix is non-singular.
Each of the $\tilde{M}^{(k)}$'s has at least $\iota - 2 \ro$ nullspace vectors of
degree at most $d$.
Therefore, in at most $q$ calls to Algorithm {\sf Nullspace minimal vectors} 
(see also Remark~\ref{rem:bigp}) on the 
$\tilde{M}^{(k)}$'s with $\delta =d$ we compute $(q-1)\ro +
(m-(q-1)\ro -2\ro) = m -2\ro$ nullspace vectors of degrees less than~$d$.
This is exactly in $q$ calls if exactly $\iota - 2 \ro$ nullspace
vectors have degree less than $d$ at each call, or if exactly 
$\iota - 2 \ro$ vectors are kept.
 Otherwise, a greedy
strategy as in previous section may need less calls.
Without giving the details here, we remark that {\em ad hoc} 
successive index choices for constructing the submatrices
$\tilde{M}^{(k)}$'s will lead 
to $m -2\ro$ linearly independent vectors (see
Proposition~\ref{prop:nullspace} and its proof).
Once this is done, we are led to a remaining $\min\{m,2\ro\} \times \ro$
matrix $\tilde{M}^{(q+1)}$ whose nullspace can be computed by Algorithm {\sf
  Nullspace}$_{2\text{\sf n}}$. If $m\leq 2\ro$ then
$\tilde{M}^{(q+1)}$ is simply the input matrix~$M$.
Again, we ensure independency by {\em ad hoc} row index choices.

We do not further detail the proof of the algorithm which relies on similar
techniques than those used for the proof of 
Proposition~\ref{prop:nullspace}. The $m-\ro$ computed vectors at
Step\,(c) and Step\,(d) are
in the nullspaces of full rank submatrices  with $\ro$ columns of $\tilde{M}$, hence are in the
nullspace of $\tilde{M}$.  The check\,(e) ensures that they are in
the nullspace of $M$.

\begin{theorem} \label{thetheo}
Let $M \in \K[x] ^{m \times n}$ be of degree $d$.
The rank $r$ of $M$ and $m-r$ linearly independent polynomial
vectors in
the nullspace of $M$ can be computed in 
\begin{equation}\label{eq:costgen}
O( nm\MM(r,d)/r^2 + (m/r +\log r)(\MM(r,d) \log (rd) + r^2 \Pgcd(d) \log r + r \PM(rd))  )
\end{equation} 
hence $\psoftO{nmr^{\omega -2}d}$ 
operations in $\K$ by a randomized Las Vegas (certified) algorithm.
The degree sum of the computed nullspace vectors is less than 
$rd \lceil \log _2 r \rceil + (m-2r)d$. 
\end{theorem}
\ifshortversion
\else
\ifshortversion
\vspace*{0.4cm}
\namedProof{Theorem~\ref{thetheo} page \pageref{thetheo}}
\else
\Proof
\fi
The cost for computing $\tilde{M}$ using Lemma~\ref{lem:Mnr} is
bounded by 
$O( nm\MM(r,d)/r^2)$. The top $\ro \times \ro$ matrix is made
non-singular by pre-multiplication by a random constant matrix $Q
\in \K ^{m \times m}$
(see Algorithm {\sf Nullspace}$_{2\text{\sf n}}$) in $O(\MM(n,d))$.
Since only the first~$\ro$ rows of~$M$ need to be modified, 
the first $\ro$ rows of $Q$ are randomly chosen in~$K$, and the last
$m-\ro$ are fixed to $[0~~I_{m-\ro}]^T$. The cost of the
multiplication by~$Q$ is $O( (m/r)(\MM(r,d)))$.
At Step\,(c) we run Algorithm {\sf Nullspace minimal vectors}
$q=O(m/r)$ times on matrices of dimensions $O(r)$.
Each call has cost~(\ref{eq:costcompress}) with $n=r$.
Then at Step\,(d) one call to Algorithm {\sf Nullspace}$_{2\text{\sf n}}$
has cost~(\ref{eq:cost2n}) with $m$ and $n$ in~$O(r)$.
The two latter costs give the factor of $O(m/r +\log r)$
in~(\ref{eq:costgen}).
The final check at Step\,(e) is done in $q+1$ multiplications using
the special form of the intermediate results of Step\,(c) and
Step\,(d).
For one output of  {\sf Nullspace minimal vectors} 
at Step\,(c), the check is done in $O(n/r)\MM(r,d)$
operations, therefore $q$ calls lead to a check in 
 $O((nm)\MM(r,d)/r^2)$.
As done in Corollary~\ref{cor:costpnullspace} for computing
$\lambda$,
the check involving the output of Algorithm {\sf
  Nullspace}$_{2\text{\sf n}}$
is done by splitting the large degrees in the $N'_i$'s, and by
forming an $(\min\{m,2\ro\}-\ro) \times m$ matrix of degree $d$,
the multiplication by~$M$ is done in O$((nm)\MM(r,d)/r^2)$
operations.

The degree bound follows from the fact that the minimal vectors 
computations of Step\,(c) lead to $m-2r$ vectors of degrees at
most~$d$.
Proposition~\ref{prop:nullspace} gives the term
$rd\lceil \log _2 r \rceil$ for the degree sum bound for Step\,(d) outputs.
\Endproof

\fi

For $m \leq 2r$ we have already commented after
Proposition~\ref{prop:nullspace}
the quality of the degree sum bound $rd \lceil \log _2 r \rceil$.
For $m \gg r$,
since the sum of the Kronecker indices is no more than $rd$, we
see that the bound we propose in Theorem~\ref{thetheo} 
is within a factor asymptotically $m/r$ from the optimal.
A more accurate ``tri-parameter'' analysis---with respect to $n$, $m$ and
 $r$---remains to be done. It may first require slight modifications of the $\sigma$-basis
 algorithm
of~\cite{BeLa94,GJV03-2} that we use for computing minimal vectors, 
and a corresponding cost analysis
especially with respect to $r$ when  $m \gg r$.

We conclude with a simplified expression of the cost for $n=m$ and using
$r \leq n$. The polynomial matrix multiplication has cost given
by~(\ref{eq:costMM1})
or~(\ref{eq:costMM2}), and we take $\PM(d)=O(d \log d \log \log d)$~\cite{CaKa91}.

\begin{corollary} 
The rank $r$ of $M \in \K[x] ^{n \times n}$ of degree $d$, 
and $m-r$ linearly independent polynomial
vectors in
the nullspace of $M$ can be computed in 
$$O(\MM(n,d)(\log^2 n + \log n \log d) + n^2\Pgcd(d)\log ^2 n \log
\log n)$$
hence $\psoftO{n^{\omega}d}$
operations in $\K$ by a randomized Las Vegas (certified) algorithm.
\end{corollary}

\begin{remark} \label{rem:genrand}
We did not detail the probability analysis. Random values in $\K$ occur
for: the choice of $P$ concerning the denominator matrix $S$ and
  the right fraction degree bound in   Proposition~\ref{prop:lrHp};
the choice of $Q$ in Lemma~\ref{lem:Q} 
for the degree dominance of the last columns in bases, 
and as linear independence conditioning in the different
algorithms;
the point $x_0$ in Algorithms {\sf Minimal
nullspace vectors} and {\sf Nullspace}$_{2\text{\sf n}}$;
the random conditioning of $M$ into $\tilde{M}$ in Lemma~\ref{lem:Mnr}.
Our algorithms are deterministic if random values are replaced by
symbolic variables. For a given input matrix $M$, the algorithm
succeeds 
if the random values do not form a
zero of a fixed polynomial over $\K$ in the latter variables.
This happens only with small probability  if the random values are chosen from
a subset of $\K$ of appropriate
cardinality~\cite{DeMLi78,Zip79,Sch80}
(see also our comments in Introduction).
\end{remark}


\section*{Concluding remarks}

We compute a
$\K (x)$-nullspace basis of an input matrix over $\K[x]$ as the
union of few minimal $\K[x]$-basis of submatrices of
the input matrix. It remains to compute a minimal basis with an
analogous complexity estimate. 
A possible direction of work here is to ensure the irreducibility 
of the output
basis either on the fly or {\em a posteriori}.

Subsequent work may also concern the applicability of our 
compression\,/\,uncompression scheme
to other problems such as questions about matrix approximants or 
block structured matrices.

Computing a nullspace basis is added to the recent list of problems
that can be solved in about the same number of operations as for
multiplying two matrix polynomials.
We hope that this will help in making progress for the characteristic
polynomial~\cite{Kal92,KaVi04-2}, and for (non-generic) matrix
inversion~\cite{JeVi04}.


\bibliographystyle{plain}
\bibliography{basis}


\end{document}